\documentclass[11pt]{article}

\usepackage{graphicx}
\usepackage{amssymb}

\renewcommand{\SS}{\scriptscriptstyle}

\title{Towards a unique formula for neutrino oscillations in vacuum}
\author{M.~Beuthe}
\date{{\it \small Institut de Physique Th\'eorique, Universit\'e catholique de
Louvain, B-1348 Louvain-la-Neuve, Belgium}\\ \small(February 9,
2002)}

\begin{document}

\maketitle
\begin{abstract}
We show that all correct results obtained by applying quantum
field theory to neutrino oscillations can be understood in terms
of a single oscillation formula. In particular, the model proposed
by Grimus and Stockinger is shown to be a subcase of the model
proposed by Giunti, Kim and Lee, while the new oscillation
formulas proposed by Ioannisian and Pilaftsis and by Shtanov are
disproved. We derive an oscillation formula without making any
relativistic assumption and taking into account the dispersion, so
that the result is valid for both neutrinos and mesons. This
unification gives a stronger phenomenological basis to the
neutrino oscillation formula. We also prove that the coherence
length can be increased without bound by more accurate energy
measurements. Finally, we insist on the wave packet interpretation
of the quantum field treatment of oscillations.
\end{abstract}

\section{Introduction}

Nowadays the study of neutrino oscillations is the particle
physicists' best hope to learn how to extend the Standard Model,
since it provides information on the neutrino masses and mixings
which are thought to originate in new physics. Neutrino
oscillations are an example of flavor oscillations, which arise
when the particles produced and detected in an experiment are
superpositions of different mass eigenstates. The interference
between the propagating mass eigenstates leads to a spatial
oscillation of the detection probability of a neutrino with a
given flavor. The standard oscillation formula, expressing the
probability to detect a neutrino of momentum $p_0$ emitted with a
flavor $\alpha$ as having a flavor $\beta$ after a propagation on
a distance $L$, reads \cite{bilenky}
\begin{equation}
  {\cal P}_{\alpha \to \beta}(L) =
   \sum_{i,j} V_{i\alpha} \, V_{i\beta}^* \, V_{j\alpha}^* \, V_{j\beta}
   \; \exp \left(
   - i \frac{m_i^2-m_j^2}{2p_0} \, L \right) \, ,
  \label{standard}
\end{equation}
where $m_{i,j}$ are the neutrino mass eigenstates and $V$ is the
matrix relating the flavor to the mass fields. This phenomenon can
explain the depletion of the solar electron-neutrino flux
\cite{krastev}, as well as the up/down asymmetry of the neutrino
flux produced by cosmic rays in the atmosphere \cite{kajita}.
Moreover, the long baseline experiment K2K hints to a
muon-neutrino depletion on a terrestrial distance \cite{K2K},
while controversial evidence for neutrino oscillations has
appeared in the laboratory experiment LSND \cite{lsnd}.

In view of the importance of neutrino oscillations, it is rather
unsettling that the neutrino oscillation formula (\ref{standard})
used to analyze experimental results is nearly always derived in a
completely inconsistent way. In particular, this derivation
usually starts with unlocalized plane wave states and ends,
strangely, with a space-dependent oscillation formula. This
hocus-pocus requires at least two arbitrary assumptions: the equal
time prescription, stating that the propagation time is the same
for the different mass eigenstates, and the classical propagation
condition, $x=vt$ (see \cite{beuthe} for a review). Additional
assumptions have generated endless discussions about the equality
of the energy-momenta of the different mass eigenstates. Moreover,
the rejection of the equal time prescription by some authors has
lead them to predict oscillation lengths differing from the
standard result by some factor (a factor 2 is often quoted).

A more consistent derivation of the neutrino oscillation formula
has been done in the {\it intermediate wave packet model}, in
which the neutrino mass eigenstates are represented by wave
packets propagating in space-time \cite{kayser}. Although the
classical propagation condition can be dispensed with, this method
implicitly requires the use of the arbitrary equal time
prescription in order to obtain the standard result for the
oscillation length. Moreover, this derivation is not wholly
consistent because it involves flavor eigenstates which cannot be
well-defined \cite{giunti92}. Finally, while the 3-momentum
uncertainty is naturally included in this model, the energy
uncertainty has to be put in by hand, with the result that there
is no agreement on whether the coherence length has an upper bound
or not \cite{kiers96, giunti98b}.

In order to solve the various problems of the intermediate wave
packet model, different models using quantum field theory have
been proposed. They can be grouped in four categories. The first
category groups the {\it external wave packet models}, the best
example of which is given by the seminal paper by Giunti, Kim, Lee
and Lee \cite{giunti93}. In these models, the neutrino is
considered as an internal line of a Feynman diagram, propagating
between a source and a detector, which are represented by in- and
outgoing wave packets \cite{giunti93,giunti98,cardall,beuthe}. A
variation on this theme is the Kiers-Weiss model, in which the
external wave packets are replaced by quantum oscillators
\cite{kiers98}. The second category groups the {\it stationary
boundary conditions models}, the best example of which is given by
the Grimus-Stockinger model \cite{grimus96}. They are very similar
to the external wave packet models, except that the wave packets
are replaced by states with a well-defined energy, i.e. stationary
states \cite{kobzarev,grimus96,grimus99,ioannisian,chung}. The
third category groups models where the neutrino is represented by
its propagator and coupled at production with a source, though not
coupled at detection with anything \cite{srivastava,shtanov}. Note
that this kind of model requires the equal time prescription, as
in the intermediate wave packet model, in order to obtain the
standard oscillation length, since the detection mechanism is left
unspecified. The fourth category includes the Blasone-Vitiello
model, in which the construction of a Fock space of flavor
eigenstates is attempted \cite{blasone}.

While few authors deny that the most rigorous treatment of
oscillations is done in a quantum field framework, the variety of
such models has not made a good case for their widespread
acceptance. Besides, these models sometimes yield conflicting
oscillation formulas, either because of their different
assumptions or because of their different approximation methods.
The main aim of this article is to show which of these results are
correct and how they can all be understood within the external
wave packet model. In particular, the Grimus-Stockinger
oscillation formula will be seen to be a limiting case of the
Giunti-Kim-Lee formula, while the Ioannisian-Pilaftsis and Shtanov
new oscillation formulas will be shown to be false. At the same
time, the role of accurate energy measurements in the coherence
length will be clarified.

Contrary to the existing literature, our calculations take into
account the dispersion and do not resort to a relativistic limit.
In this way, the final oscillation formula will be valid as well
for neutrinos as for K and B mesons (it will be shown that the
spin can be neglected). This unification of the neutrino and meson
oscillation formulas makes it much more difficult to accept
nonstandard results, since the meson oscillations are well studied
and the meson oscillation length has been cross-checked with
different methods. Another of our concerns will be to underline
the quantum mechanical interpretation of the external wave packet
model. The quantum field theory models are often perceived as
complicated and obscure, whereas they also exist in simple
versions \cite{kobzarev} and have a clear physical interpretation
in direct correspondence with intermediate wave packet models.

The outline of the article is as follows. In the second section,
the gaussian external wave packet model is reviewed and all useful
definitions are given. The flavor-mixing amplitude is expressed as
a convolution of the neutrino propagator with a function depending
on the overlap of the external wave packets. This section does not
contain new material and follows the notation of \cite{giunti98}.
In the third section, a new way to evaluate the convolution is
explained, leading to an explicit oscillation formula. It is then
shown that the coherence length, beyond which oscillations vanish,
can be increased without bound by more accurate energy
measurements. As a corollary, it is proved that the stationary
boundary conditions model is a special case of the external wave
packet model. In the fourth section, another new method to
evaluate the convolution is presented, yielding the explicit
dependence of the amplitude on time and distance. This method
allows a clear physical interpretation of the amplitude in terms
of wave packets. In particular, the propagation range is
subdivided into three regimes by two thresholds marking the onset
of the transversal and longitudinal dispersion of the wave packet.
With this method, the new oscillation formulas proposed in
Refs.~\cite{ioannisian} and \cite{shtanov} can be disproved.
Finally, we show that the Blasone-Vitiello oscillation formula
\cite{blasone} is phenomenologically equivalent to the standard
result.

\section{The external wave packet model}

The propagating process of a particle between a source and a
detector (indicated by dotted circles) is symbolized by the
following diagram:
\begin{center}
\includegraphics[width=12cm]{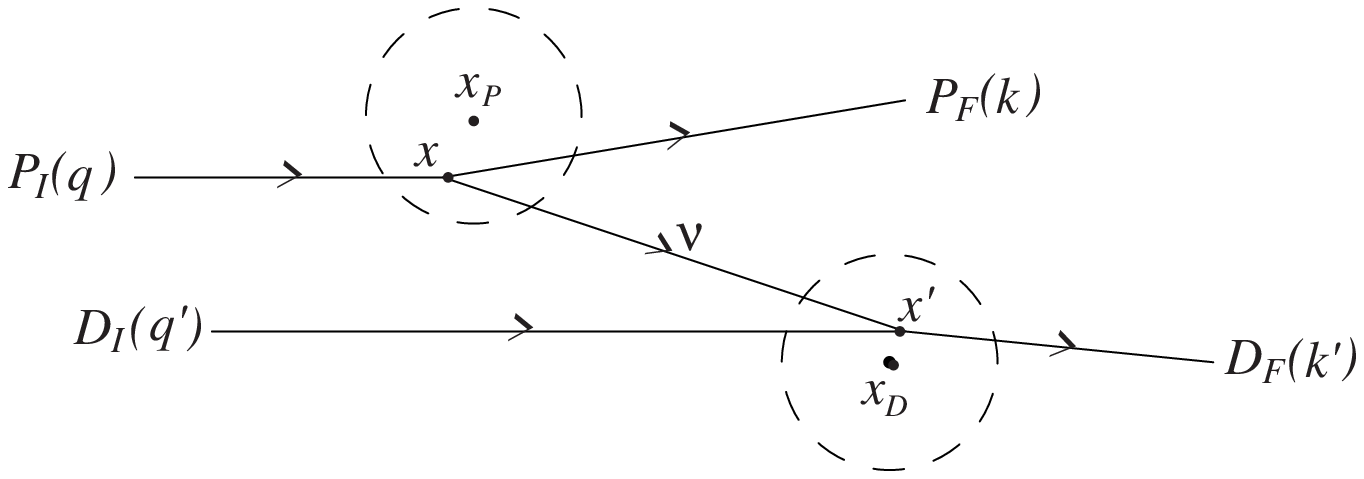}

FIG.\ 1. Propagation of a particle $\nu$ between a source and a
detector, centered in $x_{\SS P}$ and $x_{\SS D}$.
\end{center}

The arrows indicate the momentum flow. $P_I$ represents the set of
incoming particles, of total momentum $q$, arriving in the
production region (or source), which is centered around the point
$( t_{\SS P},{\bf x}_{\SS P} )$. $P_F$ represents the set of
outgoing particles, of total momentum $k$, coming from the
production region, with the exception of the intermediate neutrino
$\nu$ whose propagation is studied. $D_I$, $D_F$ and $( t_{\SS
D},{\bf x}_{\SS D} )$ are defined similarly, but apply to the
detection process. The interaction points at production and
detection are noted $x$ and $x'$, respectively. All particles are
assumed to be stable. The internal line could also represent an
antineutrino, but we shall assume that the experimental conditions
are such that a quasi-real neutrino propagates on a macroscopic
distance, transferring positive energy from $x$ to $x'$. If
$x'^0-x^0$ is a macroscopic time, it will be seen that the
neutrino $\nu$ contributes to the propagation, but not its
antiparticle.

Since known neutrinos are much lighter than the experimental
energy thresholds, their masses can be neglected in the
interaction vertices and in the numerator of the neutrino
propagator. In that case, the spin structure factorizes from the
sum over the mass eigenstates, so that the computation of the
oscillation formula can be carried out as if the neutrinos were
scalar. If one wants to consider nonrelativistic neutrinos,
factorization also occurs for nearly degenerate masses. The only
case where factorization does not occur is nonrelativistic
neutrinos with very different masses. However, spatial
oscillations vanish in these conditions since it is experimentally
possible to determine which mass eigenstate contributes to the
amplitude. In that case, a treatment with wave packets is not
necessary, since the flavor-changing probability does not depend
on the propagation distance. In the following we shall assume that
the neutrinos are either relativistic or nearly degenerate in
mass. Thus the spin structure can be absorbed in an overall
multiplying factor.

The external wave packets are expressed in terms of their
momentum-space wave function: $|\psi > = \int [d{\bf k}] \,
\psi({\bf k},{\bf K}) \, |{\bf k} >$, where $[d{\bf k}]$ is the
integration measure in 3-momentum space and ${\bf K}$ is the
average 3-momentum. If $\psi({\bf k},{\bf K})$ is the momentum
space wave function of a wave packet centered in ${\bf x}={\bf 0}$
at time $t=0$, then
$$
   \Psi({\bf k},{\bf K},{\bf x}_0,t_0)
   = \psi({\bf k},{\bf K}) \,
   e^{ i E({\bf k}) t_0 - i \, {\bf k} \cdot {\bf x}_0 } \, ,
$$
is the momentum-space wave function of a wave packet centered in
${\bf x}_0$ at time $t_0$.

Without loss of generality, let us choose to work with only one
particle in $P_I(q)$, in $P_F(k)$, in $D_I(q')$ and in $D_F(k')$.
The extension to a larger number is straightforward and would only
complicate the notation. The wave packets are built such that
those involved in the production of the $\nu$ are centered at
${\bf x}_{\SS P}$ at time $t_{\SS P}$, while those involved in the
detection of the $\nu$ are centered at ${\bf x}_{\SS D}$ at time
$t_{\SS D}$. They are noted
\begin{eqnarray*}
   |\, P_I \!> &=& \int [d{\bf q}] \,
   \Psi_{ P_{\SS I} } \left( {\bf q},{\bf Q},{\bf x}_{\SS P},t_{\SS P} \right)
   |\, P_{\SS I} ({\bf q}) \!>
   \\
   |\, P_F \!> &=& \int [d{\bf k}] \,
   \Psi_{ P_{\SS F} } \left( {\bf k},{\bf K},{\bf x}_{\SS P},t_{\SS P} \right)
   |\, P_{\SS F} ({\bf k}) \!>
   \\
   |\, D_I \!> &=& \int [d{\bf q'}] \,
   \Psi_{ D_{\SS I} } \left( {\bf q'},{\bf Q'},{\bf x}_{\SS D},t_{\SS D} \right)
   |\, D_{\SS I} ({\bf q'}) \!>
   \\
   |\, D_F \!> &=& \int [d{\bf k'}] \,
   \Psi_{ D_{\SS F} } \left( {\bf k'},{\bf K'},{\bf x}_{\SS D},t_{\SS D} \right)
   |\, D_{\SS F} ({\bf k'}) \!> \, .
\end{eqnarray*}

Let us first suppose that there is only one mass eigenstate $m_j$
associated to the propagating neutrino of FIG.\ 1. The general
formula of the connected amplitude corresponding to this process
is given by
$$
   {\cal A}_j =
   <\! P_F,D_F \,| \,
   \hat T \left( \exp \left( - i \int d^4x \, {\cal H}_I \right)  \right)
   - {\bf 1}
   |\, P_I,D_I \!> \, ,
$$
where ${\cal H}_I$ is the interaction Lagrangian for the neutrino
and $\hat T$ is the time ordering operator. Let $g$ be the
coupling constant of $\nu$ with the other fields. Expanding the
amplitude to order $g^2$, and inserting the wave packets
expressions, one obtains \cite{jacob}
\begin{equation}
   {\cal A}_j = i \int \frac{d^4p}{(2\pi)^4} \,
   \frac{ \psi(p^0,{\bf p}) }{ p^2-m_j^2+i\epsilon} \,
    e^{-i p^0 T + i \, {\bf p} \cdot {\bf L} } \, ,
  \label{ampli}
\end{equation}
where the average propagation time $T$ is defined by $T=x_{\SS
D}^0-x_{\SS P}^0$ and the average propagation distance by ${\bf
L}={\bf x}_{\SS D}-{\bf x}_{\SS P}$.

The {\it overlap function} $\psi(p^0,{\bf p})$ represents the
overlap of the incoming and outgoing wave packets, both at the
source and at the detector. It is defined by
\begin{eqnarray}
   \psi(p^0,{\bf p}) =
   \int d^4x \, e^{ipx} \int d^4x' \,e^{-ipx'}
   \int [d{\bf q}]  \, \psi_{\SS P_{in}}   ({\bf q},{\bf Q}) \, e^{-iqx}
   \int [d{\bf k}]  \, \psi_{\SS P_{out}}^* ({\bf k},{\bf K}) \, e^{ikx}
   \nonumber \\  \times \
   \int [d{\bf q'}] \, \psi_{\SS D_{in}}   ({\bf q'},{\bf Q'}) \, e^{-iq'x'}
   \int [d{\bf k'}] \, \psi_{\SS D_{out}}^* ({\bf k'},{\bf K'}) \, e^{ik'x'} \,
   M_P(q,k) \, M_D(q',k') \, ,
   \label{overlap}
\end{eqnarray}
where $M_P(q,k)$ and $M_D(q',k')$ are the interaction amplitudes
at production and detection. All external particles are on-shell,
i.e.\ $
   q^0 = E_{ P_{\SS I}}({\bf q}) = \sqrt{ {\bf q}^2 + m^2_{P_{\SS I} } } \, ,
$ and so on. Note that the overlap function is independent of
$x_{\SS P}$ and $x_{\SS D}$. The integrals over $x$ and $x'$ in
Eq.~(\ref{overlap}) yield delta functions, which impose
energy-momentum conservation at the source and the detector. The
amplitude corresponding to the propagation of an antineutrino from
$x_{\SS P}$ to $x_{\SS D}$ can be obtained by the substitution
$p\to -p$.

Let us now consider the case where the intermediate state in FIG.\
1 is a superposition of different mass eigenstates. The fields in
the flavor and mass bases are related by a unitary matrix $V$
\cite{bilenky}
$$
   \nu_\alpha = \sum_j \, V^\dagger_{\alpha j} \, \nu_j \, ,
$$
where Greek and Latin indices refer to the flavor and to the mass
basis, respectively. We assume that it is possible to tag the
initial flavor $\alpha$ and the final flavor $\beta$ of the
neutrino. The flavor-mixing amplitude ${\cal
A}_{\alpha\to\beta}(T,{\bf L})$ is defined as the transition
amplitude corresponding to the propagation over a time $T$ and a
distance ${\bf L}$ of a neutrino initially of flavor $\alpha$, but
detected with a flavor $\beta$.

The flavor-mixing amplitude can be expressed as a linear
combination of amplitudes ${\cal A}_j$ corresponding to the
propagation of different mass eigenstates:
\begin{equation}
   {\cal A}_{\alpha\to\beta}(T,{\bf L})
   = \sum_j V_{\beta j}^\dagger \, {\cal A}_j \, V_{j\alpha} \, ,
   \label{amplitot}
\end{equation}
where ${\cal A}_j$ is given by Eq.~(\ref{ampli}).

As the propagation time is not measured in experiments, the
transition probability has to be averaged over $T$:
\begin{equation}
   {\cal P}_{\alpha \to \beta}({\bf L}) \sim
   \sum_{i,j} V_{i\alpha} \, V_{i\beta}^* \, V_{j\alpha}^* \, V_{j\beta} \,
   \int dT \, {\cal A}_i \, {\cal A}_j^* \, .
   \label{proba}
\end{equation}
A discussion of the proportionality factor can be found in
Ref.~\cite{cardall}.

The next task consists in performing the momentum integration in
the amplitude (\ref{ampli}). This cannot be done without doing
some assumptions on the overlap function (\ref{overlap}), which
should be as weak as possible. A fairly general oscillation
formula can be obtained by approximating the in- and outgoing
particles with gaussian wave packets \cite{giunti93,giunti98}. A
gaussian wave packet is defined by
$$
   \psi_{\SS P_{in}}({\bf q},{\bf Q}) =
   \left( \frac{2\pi}{ \sigma_{p\SS P_{in}}^2 } \right)^{3/4}
   \exp \left( - \frac{({\bf q}-{\bf Q})^2}{ 4\sigma_{p\SS P_{in}}^2 }
        \right) \, ,
$$
where $\sigma_{p\SS P_{in}}$ is the width of the wave packet in
momentum space. It is also useful to define the width
$\sigma_{x\SS P_{in}}$ in configuration space by $\sigma_{p\SS
P_{in}}\sigma_{x\SS P_{in}} = 1/2$. The normalization of the
states is defined as in Ref.~\cite{peskin}, so that
$\int\frac{d{\bf q}}{(2\pi)^3}|\psi({\bf q})|^2=1$.

If the wave packet is sharply peaked around its average momentum
${\bf Q}$, i.e.\ $\sigma_{\SS P_{in}} \ll E_{\SS P_{in}}$, the
energy can be approximated by
$$
   E_{\SS P_{in}}({\bf q}) \cong E_{\SS P_{in}}({\bf Q})
   + {\bf v}_{\SS P_{in}} \cdot ({\bf q}-{\bf Q}) \, ,
$$
where $E_{ P_{in} }({\bf q}) = \sqrt{{\bf q}^2 + m_{\SS
P_{in}}^2}$ and ${\bf v}_{\SS P_{in}} = {\bf Q}/E_{\SS
P_{in}}({\bf Q})$. This approximation means that dispersion is
neglected in the external wave packets. The factors $M_P(q,k)$ and
$M_D(q',k')$ multiplying the exponential vary slowly over the
width of the wave packet and can be approximated by their value at
the average momentum. They can be factorized outside the sum over
the mass eigenstates since we have assumed that the neutrinos are
either relativistic or nearly degenerate in mass.

With these approximations, the momentum integrations in the
overlap function (\ref{overlap}) can be done analytically. One
then performs the integrations on ${\bf x}$, ${\bf x}'$ and $t$,
$t'$, which are all gaussian. The final result can be written as
\begin{equation}
   \psi(p^0,{\bf p}) = N \,
   \exp \left( -f_{\SS P}(p^0,{\bf p}) - f_{\SS D}(p^0,{\bf p}) \right) \, ,
   \label{overlapPD}
\end{equation}
with
$$
   f_{\SS P}(p^0,{\bf p}) =
   \frac{ \left( {\bf p}-{\bf p}_{\SS P} \right)^2}{4\sigma_{p \SS P}^2}
   +\frac{ \left(
   p^0 - E_{\SS P} -({\bf p} - {\bf p}_{\SS P}) \cdot {\bf v}_{\SS P}\right)^2}
         {4\sigma_{e\SS P}^2} \, ,
$$
where $E_{\SS P}=E_{\SS P_{in}} - E_{\SS P_{out}}$, ${\bf p}_{\SS
P}={\bf Q} - {\bf K}$. The function $f_{\SS D}(p^0,{\bf p})$ is
defined in the same way, with the index $P$ replaced by $D$,
except for the energy-momentum which is defined so as to be
positive: $E_{\SS D}=E_{\SS D_{out}} - E_{\SS D_{in}}$ and ${\bf
p}_{\SS D}={\bf K}' - {\bf Q}'$. The constant $N$ includes the
normalization constants as well as the factors $M_{P,D}$ evaluated
at the average momenta.

A new width $\sigma_{p\SS P}$ has been defined by $\sigma_{p\SS
P}\sigma_{x\SS P} = 1/2$, with
$$
   \frac{1}{\sigma_{x\SS P}^2} = \frac{1}{\sigma_{x\SS P_{in}}^2}
   + \frac{1}{\sigma_{x\SS P_{out}}^2} \, .
$$
$\sigma_{p\SS P}$ can be interpreted as the {\it momentum
uncertainty at the source}. The spatial width $\sigma_{x\SS P}$ is
mainly determined by the external particle with the smallest space
width. This is expected since the production region depends on the
overlap in space-time of the external wave packets.

The symbol ${\bf v}_{\SS P}$ is defined by
$$
   {\bf v}_{\SS P} = \sigma_{x\SS P}^2 \left(
   \frac{ {\bf v}_{\SS P_{in}} }{\sigma_{x\SS P_{in}}^2} +
   \frac{ {\bf v}_{\SS P_{out}} }{\sigma_{x\SS P_{out}}^2}
   \right) \, .
$$
It can be interpreted as the velocity of the production region,
approximately equal to the velocity of the particle with the
smallest spatial spread.

The symbol $\Sigma_{\SS P}$, satisfying $0\leq\Sigma_{\SS P}\leq
1$, is defined by
$$
   \Sigma_{\SS P} = \sigma_{x\SS P}^2 \left(
   \frac{ {\bf v}_{\SS P_{in}}^2 }{\sigma_{x\SS P_{in}}^2} +
   \frac{ {\bf v}_{\SS P_{out}}^2 }{\sigma_{x\SS P_{out}}^2}
   \right) \, .
$$

Finally, the quantity
$$
   \sigma_{e\SS P}^2 = \sigma_{p \SS P}^2
   \left( \Sigma_{\SS P} - {\bf v}_{\SS P}^2 \right)
   \leq \sigma_{p \SS P}^2
$$
can be interpreted as the {\it energy uncertainty at the source},
or also as the inverse of the time of overlap of wave packets
during the production process. Indeed, one can show that
\begin{equation}
   \sigma_{e\SS P}^2 = \sum_{\alpha<\beta}
   \frac{ \sigma_{x\SS P}^2 }{ 4\sigma_{x\alpha}^2 \sigma_{x\beta}^2 } \,
   \left( {\bf v}_\alpha - {\bf v}_\beta \right)^2 \, ,
   \label{energywidth}
\end{equation}
where the sum is over all wave packets involved in the production
process. This sum is dominated by the term including the two
smallest wave packets in configuration space (unless their
velocities are nearly equal). If $\sigma_{x1}$ is the smallest
width and $\sigma_{x2}$ the second smallest, one obtains
\begin{equation}
   \sigma_{e\SS P} \sim \frac{ |{\bf v}_1 - {\bf v}_2| }{ \sigma_{x2} }
   \sim \frac{1}{T^{overlap}_{\SS P}} \, ,
   \label{timeoverlap}
\end{equation}
where $T^{overlap}_{\SS P}$ is defined as the duration of the
production process. Thus, $\sigma_{e\SS P}$ can be interpreted as
the energy uncertainty at the source, since it is proportional to
the inverse of the time of overlap of the external wave packets at
the source. The quantities $\sigma_{x\SS D}$, $\sigma_{p\SS D}$,
${\bf v}_{\SS D}$, $\Sigma_{\SS D}$, $\sigma_{e\SS D}$,
$T^{overlap}_{\SS D}$ have similar definitions and properties.

Note that stationary boundary conditions are recovered by setting
${\bf v}_{\SS P,D}=0$ and $\sigma_{e\SS P,D}=0$, with
$\sigma_{p\SS P,D}$ different from zero. Moreover, it is
reasonable to impose the constraints $|{\bf v}_{\SS
P,D}|T^{overlap}_{\SS P,D}\lesssim S_{\SS P,D}$, where $S_{\SS P}$
(resp. $S_{\SS D}$) is the size of the macroscopic region of
production (resp. detection). Thus we shall assume the following
constraints:
\begin{equation}
  \frac{|{\bf v}_{\SS P,D}|}{\sigma_{e\SS P,D}}
  \lesssim S_{\SS P,D} \, .
  \label{statlim}
\end{equation}
These bounds are very conservative, since we shall see that
stationary models such as those found in
Refs.~\cite{grimus96,ioannisian} are recovered by setting $|{\bf
v}_{\SS P,D}|/\sigma_{e\SS P,D}=0$. In the example of the
Grimus-Stockinger model \cite{grimus96}, an initial stationary
neutron (${\bf v}_{\SS P_{in},n}=0$) decays into a stationary
proton (${\bf v}_{\SS P_{out},pr}=0$), a `plane-wave' electron
($\sigma_{x\SS P_{out},el}=\infty$) and the intermediate neutrino.
At detection, the neutrino collides with a stationary electron
(${\bf v}_{\SS D_{in}}=0$) and the outgoing neutrino and electron
are represented as plane waves ($\sigma_{x\SS
D_{out},\nu}=\sigma_{x\SS D_{out},el}=\infty$).

As the propagation distance is macroscopic, only processes
satisfying global conservation of energy-momentum have a
non-negligible probability of occurring. Since our aim is not to
prove this well-known fact, we impose that
\begin{equation}
   {\bf p}_{\SS P} = {\bf p}_{\SS D} \equiv {\bf p}_0
   \hspace{1cm} \mbox{and} \hspace{1cm}
   E_{\SS P} = E_{\SS D} \equiv E_0 \, .
   \label{conserv}
\end{equation}
This approximation allows to do expansions around ${\bf p}_0$ and
$E_0$. An associated velocity can be defined by ${\bf v}_0={\bf
p}_0/E_0$.

\section{The oscillation formula, with or without stationary boundary conditions}
\label{section3}

In this section, we shall explain a quick method to evaluate the
flavor-changing probability (\ref{proba}) with the amplitude given
by Eq.~(\ref{ampli}) and the overlap function by
Eq.~(\ref{overlapPD}). Since the experimental conditions are such
that the propagating particle is on-shell, the main contribution
to the transition amplitude (\ref{ampli}) comes from the pole of
the propagator. However, one has to be careful with the choice of
the contour as the analytic continuation of most overlap functions
diverges at infinity in the complex plane. The integration on the
3-momentum can be done with the help of the Grimus-Stockinger
theorem \cite{grimus96}. Let $\psi({\bf p})$ be a 3 times
continuously differentiable function on ${\bf R}^3$ such that
$\psi$ itself and all its first and second derivatives decrease at
least like $1/{\bf p}^2$ for $|{\bf p}| \to\infty$. Then, for a
real number $A>0$,
$$
   \int d^3p \,
   \frac{\psi({\bf p}) \, e^{i \, \bf p \cdot L}}{A - {\bf p}^2 + i \epsilon}
   \; \stackrel{ L \to \infty }{\longrightarrow} \;
   - \frac{2\pi^2}{L} \, \psi(\sqrt{\!A}\,{\bf l}) \, e^{i\sqrt{\!A}L} \, ,
$$
where $L=|{\bf L}|$ and ${\bf l}={\bf L}/L$. For $A<0$, the
integral decreases like $L^{-2}$.

The remaining energy integral in the amplitude (\ref{ampli}) can
be done by a saddle-point approximation \cite{giunti98}. However,
it is quicker to perform first the time average in the probability
(\ref{proba}), which yields a delta function, and makes one of the
energy integrations trivial. At this stage, one has
\begin{equation}
   \int dT \, {\cal A}_i {\cal A}_j^* \sim \frac{1}{L^2} \,
   \int dE \, \psi(E,q_i\,{\bf l}) \,
   \psi^*(E,q_j\,{\bf l}) \,
   e^{ i (q_i-q_j)L } \, ,
   \label{incohsumQFT}
\end{equation}
where $\psi(E,{\bf p})$ is the overlap function defined by
Eq.~(\ref{overlapPD}) and $q_j=\sqrt{E^2-m_j^2}$.

Eq.~(\ref{incohsumQFT}) shows that the transition probability can
be interpreted as an incoherent sum (i.e.\ occurring in the
probability) over energy eigenstates: interference occurs only
between the components of $\psi(E,{\bf p})$ having the same energy
\cite{sudarsky}. In this way, the correspondence between models
with and without stationary boundary conditions is obvious: {\it
the time-integrated nonstationary probability is equivalent to the
energy-integrated stationary probability}. For example, the
oscillation formula obtained by Grimus and Stockinger with
stationary boundary conditions \cite{grimus96} has the form of the
integrand in the right-hand side of Eq.~(\ref{incohsumQFT}). This
equivalence confirms that the stationary case can be obtained from
the more general nonstationary case in the limit of a vanishing
energy width. This limit will have to be checked explicitly on the
final oscillation formula, as it is not obvious that the
approximations involved in the computations respect this feature.

If the coordinate system is chosen so that ${\bf L}$ is oriented
along a coordinate axis, it is easy to rewrite the integral
(\ref{incohsumQFT}) as
\begin{equation}
   \int dT \, {\cal A}_i {\cal A}_j^* \sim \frac{g^2({\bf l})}{L^2} \,
   \int dE \;
   e^{ i (q_i-q_j)L - f_i(E) - f_j(E) } \, ,
   \label{timeaverageshort}
\end{equation}
with the definitions $f_j(E)=f_{j\SS P}(E) + f_{j\SS D}(E)$ and
\begin{equation}
   f_{j\SS P}(E)=
   \frac{ \left( \sqrt{E^2-m_j^2}-p_0 \right)^2}{4\sigma_{p \SS P}^2}
   +\frac{ \left( E - E_0 - \left( \sqrt{E^2-m_j^2} - p_0 \right)
           v_{\SS P} \right)^2}{ 4\sigma_{e\SS P}^2 } \, ,
   \label{definitionfjPshort}
\end{equation}
where $v_{\SS P}$ and $p_0$ are the components of ${\bf v}_{\SS
P}$ and ${\bf p}_0$ along ${\bf L}$, while $E_0$ has been
redefined so as to absorb the transversal part of ${\bf p}_0\cdot
{\bf v}_{\SS P}$. The definition of $f_{j\SS D}(E)$ is similar.
The function $g({\bf l})$ expresses the geometrical constraint
between the direction of observation ${\bf l}={\bf L}/L$ and the
momentum ${\bf p}_0$:
\begin{equation}
   g({\bf l})
   = \exp \left(
                 - \frac{({\bf p}_0\times{\bf l})^2}{4\sigma_p^2}
          \right) \, ,
   \label{geom}
\end{equation}
and restricts the neutrino propagation to a cone of axis ${\bf
p}_0$ and angle $arcsin(\sigma_p/p_0)$. The momentum width
$\sigma_p$, defined by
\begin{equation}
   \frac{1}{\sigma_p^2}=
   \frac{1}{\sigma_{p\SS P}^2} + \frac{1}{\sigma_{p\SS D}^2} \, ,
   \label{definitionsigmap}
\end{equation}
is approximately equal to the smallest width among the production
and detection momentum widths. The associated width $\sigma_x$ in
configuration space is defined by $\sigma_p\sigma_x=1/2$.

The remaining energy integral in Eq.~(\ref{timeaverageshort}) can
be performed as a gaussian integral by expanding the integrand to
second order around the maximum of its modulus (this is called
Laplace's method \cite{erdelyi}). The result takes a much simpler
form if an expansion in small mass differences is done around
($E_0$, $p_0$, $m_0$, $v_0$), where $m_0^2=E_0^2-p_0^2$ and
$v_0=p_0/E_0$. The expansion parameter is noted $\epsilon$ and
refers collectively to all $\delta m_j^2/2E_0^2$, where $\delta
m_j^2=m_j^2-m_0^2$.  We shall calculate the transition probability
to ${\cal O}(\epsilon^2)$ in the real part of the argument of the
exponential (since the order ${\cal O}(\epsilon)$ vanishes) and to
${\cal O}(\epsilon)$ in the phase. The gaussian integration will
be consistent with the $\epsilon$-expansion if the extremum is
computed to ${\cal O}(\epsilon)$, the real part of the argument of
the exponential to ${\cal O}(\epsilon^2)$, the phase to ${\cal
O}(\epsilon)$, the first derivatives to ${\cal O}(\epsilon)$ and
the second derivatives to ${\cal O}(\epsilon^0)$. It is important
to take into account all these terms, if the result is to coincide
with the results obtained by the methods explained in the fourth
section.

The modulus of the integrand in Eq.~(\ref{timeaverageshort}) is
maximal for
$$
   E_{ij} = E_0 \,+\, \rho \, \frac{\delta m_i^2 + \delta m_j^2}{4 E_0}
   \;+\;  {\cal O}(\epsilon^2) \, ,
$$
where the dimensionless number $\rho$ is defined by
\begin{equation}
   \rho = \sigma_{peff}^2
   \left(
   \frac{1}{\sigma_p^2}
   - \frac{v_{\SS P}(v_0-v_{\SS P})}{\sigma_{e\SS P}^2}
   - \frac{v_{\SS D}(v_0-v_{\SS D})}{\sigma_{e\SS D}^2}
   \right) \, .
   \label{definitionrho}
\end{equation}
The {\it effective width} in momentum space $\sigma_{peff}$ is
defined by
\begin{equation}
   \frac{1}{\sigma_{peff}^2} = \frac{1}{\sigma_p^2}
   + \frac{(v_0-v_{\SS P})^2}{\sigma_{e\SS P}^2}
   + \frac{(v_0-v_{\SS D})^2}{\sigma_{e\SS D}^2} \, ,
   \label{sigmaeff}
\end{equation}
with $\sigma_p$ defined by Eq.~(\ref{definitionsigmap}). The
effective width can be interpreted as the energy-momentum width of
the oscillating particle, since it is the width of the overlap
function. Indeed, the value to ${\cal O}(\epsilon^0)$ of the
second derivative of $f_i(E) + f_j(E)$ at the extremum reads
\begin{equation}
   \frac{1}{2} \frac{d^2(f_i+f_j)}{dE^2}(E_{ij}) =
   \frac{1}{2 v_0^2 \sigma_{peff}^2} \, .
   \label{fijshortprime}
\end{equation}
The effective width is dominated by the smallest among the energy
uncertainties (recall that $\sigma_{e\SS P,D}\leq \sigma_{p\SS
P,D}$). The effective width $\sigma_{xeff}$ in configuration
space, defined by $\sigma_{peff}\sigma_{xeff}=1/2$, is then
approximately equal either to the production or to the detection
time uncertainty, depending on which one is the largest. In the
stationary limit (${\bf v}_{\SS P,D}=0$ and $\sigma_{e\SS
P,D}=0$), $\sigma_{peff}$ goes to zero.

The parameter $\rho$ has been defined so as to be in
correspondence with the notation of Ref.~\cite{giunti98}. The
symbol $\omega$ appearing in that article is related to our
notation by $\omega=\sigma_p^2/\sigma_{peff}^2$. Note that the
authors of Ref.~\cite{giunti98} do not compute $\rho$ explicitly
and also take the relativistic limit $v_0=1$. The explicit value
of $\rho$ is very interesting to know, since $\rho=0$ in the case
of stationary boundary conditions, in which case all mass
eigenstates have the same energy $E_0$.

The value to ${\cal O}(\epsilon^2)$ of $f_i(E) + f_j(E)$ at the
extremum reads
\begin{equation}
   f_i(E_{ij}) + f_j(E_{ij}) =
   \frac{(\delta m_i^2)^2+(\delta m_j^2)^2}{16 \sigma_m^2 E_0^2}
   + 2 \pi^2 \left( \frac{\rho \, \sigma_{xeff}}{L^{osc}_{ij}} \right)^2 \, ,
   \label{overlapextremum}
\end{equation}
where the {\it mass width} $\sigma_m$ is defined by
\begin{equation}
   \frac{1}{\sigma_m^2} = \sigma_{peff}^2
   \left(
   \frac{1}{\sigma_p^2} \,
   \left( \frac{1}{\sigma_{e\SS P}^2} + \frac{1}{\sigma_{e\SS D}^2} \right)
   + \frac{(v_{\SS P}-v_{\SS D})^2}{\sigma_{e\SS P}^2\sigma_{e\SS D}^2}
   \right) \, .
   \label{sigmam}
\end{equation}

The expansion of the phase present in Eq.~(\ref{timeaverageshort})
around the extremum reads
\begin{equation}
   (q_i-q_j)L \cong - \frac{\delta m_{ij}^2L}{2p_0} \,
   +  \, \frac{\delta m_{ij}^2L}{2p_0^2v_0} \, (E-E_{ij}) \, .
   \label{shortcutphase}
\end{equation}
The second derivative of the phase is of ${\cal O}(\epsilon)$ and
can be neglected with respect to the second derivative of
$f_i+f_j$.

The gaussian integration in Eq.~(\ref{timeaverageshort}) may now
be performed with the help of Eqs.~(\ref{fijshortprime}),
(\ref{overlapextremum}) and (\ref{shortcutphase}). As a result,
the flavor-mixing transition probability (\ref{proba})
corresponding to the propagation over a distance ${\bf L}=L\,{\bf
l}$ of a neutrino, initially of flavor $\alpha$, but detected with
a flavor $\beta$, reads
\begin{eqnarray}
   &&{\cal P}_{\alpha \to \beta}(L{\bf l}) \sim
   \frac{1}{L^2} \,
   \exp \left(
                 - \frac{({\bf p}_0\times{\bf l})^2}{2\sigma_p^2}
        \right) \,
   \sum_{i,j} V_{i\alpha} \, V_{i\beta}^* \, V_{j\alpha}^* \, V_{j\beta} \,
   \nonumber \\
   && \hspace{1.5cm} \times
   \exp \left(
   - 2 \pi i \frac{L}{L^{osc}_{ij}}
   - \left( \frac{L}{L^{coh}_{ij}} \right)^2
   - 2 \pi^2 \left( \frac{\rho \, \sigma_{xeff}}{L^{osc}_{ij}} \right)^2
   - \frac{(\delta m_i^2)^2+(\delta m_j^2)^2}{16 \sigma_m^2 E_0^2}
        \right) \, .
   \label{oscformula}
\end{eqnarray}
The oscillation length $L^{osc}_{ij}$ for the masses $m_i$ and
$m_j$ is given by
\begin{equation}
   L^{osc}_{ij} = \frac{4\pi p_0}{\delta m_{ij}^2} \, ,
   \label{osclength}
\end{equation}
where $\delta m_{ij}^2=m_i^2-m_j^2$ is taken to be positive. The
coherence length $L^{coh}_{ij}$ is defined by
\begin{equation}
   L^{coh}_{ij} = \frac{1}{\sqrt{2}\pi} \, \frac{p_0}{\sigma_{peff}}
                  L^{osc}_{ij} \, .
   \label{cohlength}
\end{equation}

Let us now proceed to the analysis of the oscillation formula
(\ref{oscformula}). The first term in the second exponential of
Eq.~(\ref{oscformula}) is the standard oscillation phase
proportional to the propagation distance. The second term, or {\it
coherence-length term}, leads to the vanishing of oscillations
beyond the {\it coherence length} $L^{coh}_{ij}$. This phenomenon
is partly due to the progressive separation of mass-eigenstates
wave packets propagating in space (see discussion after
Eq.~(\ref{longdisp})). The third term is a {\it localization
term}, i.e. a constraint imposing that oscillations vanish unless
the oscillation length is larger than the space-time uncertainty:
\begin{equation}
  L^{osc}_{ij} \gtrsim \rho\sigma_{xeff} \, .
  \label{local1}
\end{equation}
This condition can be rewritten as $\delta
m_{ij}^2/p_0\lesssim\sigma_{peff}/\rho$, stating that oscillations
vanish if the energy-momentum measurements allow to distinguish
between the different mass eigenstates. The coherence length and
localization conditions were already predicted in intermediate
wave packet models of neutrino oscillations
\cite{kayser,nussinov}.

The fourth term in the second exponential of
Eq.~(\ref{oscformula}) could be a matter of concern since it does
not vanish in the limit $m_i=m_j$. First of all, note that this
kind of term is not specific to our computation. For example, it
would appear in the oscillation formula (26) of
Ref.~\cite{giunti98} if the terms $S_a(E_a)$ present in Eq.~(22)
of that article had been expanded beyond zeroth order in
$m_a^2/E_a^2$. It is of interest to rewrite the term under
discussion as
\begin{equation}
  \frac{(\delta m_i^2)^2+(\delta m_j^2)^2}{16\sigma_m^2E_0^2}=
  \frac{(\delta m_{ij}^2)^2}{32\sigma_m^2E_0^2}+
  \frac{(\delta m_i^2+\delta m_j^2)^2}{32\sigma_m^2E_0^2} \, .
  \label{decomp}
\end{equation}
The first term in Eq.~(\ref{decomp}) is recognized as a
localization constraint and can be rewritten as
\begin{equation}
  L^{osc}_{ij} \gtrsim \sigma_x \, ,
  \label{local2}
\end{equation}
as $\sigma_m\sim v_0\sigma_p$ whether the stationary limit is
taken or not. The second term in Eq.~(\ref{decomp}) imposes that
\begin{equation}
  \frac{|m_i^2+m_j^2-2m_0^2|}{E_0} \lesssim \sigma_m \, .
  \label{onshell}
\end{equation}
As $m_0$ is related to the average in- and outgoing momentum by
$m_0^2=E_0^2-p_0^2$, condition (\ref{onshell}) means that the mass
eigenstates have to be on-shell with respect to $(E_0,p_0)$ within
the uncertainty $\sigma_m$. For example, this constraint is
impossible to satisfy in the mixing of relativistic and
nonrelativistic neutrinos if the available energy-momentum is such
that only the lightest neutrino can be produced. However it has no
effect on the oscillations in the two cases considered in the
present article, namely relativistic neutrinos or nearly mass
degenerate neutrinos. Condition (\ref{onshell}) should simply be
considered as expressing the conservation of energy-momentum. Such
kinematical constraints are usually not included in the
oscillation formula, though they rightly belong to it. A complete
computation of the transition probability should not only include
this energy-momentum constraint, but also similar terms arising
from the interaction amplitudes $M_{P,D}$, from the prefactor
resulting from the gaussian integration and from the spin
structure of the propagator. Actually the neutrino masses should
be expected to appear not only through mass differences but also
through their absolute values.

The oscillation formula (\ref{oscformula}) is similar to Eq.~(26)
of Ref.~\cite{giunti98}. However, three interesting new elements
appear in Eq.~(\ref{oscformula}): the parameter $\rho$ has been
computed explicitly, the oscillation formula (\ref{oscformula}) is
valid for relativistic or nonrelativistic (but nearly mass
degenerate) neutrinos, and the dispersion of the `wave packet'
corresponding to the neutrino has been taken into account. This
last fact is not at all obvious from the above computation, in
which the amplitude ${\cal A}_j$ has been inserted into the
probability before doing the integration over the energy. Had this
energy integration been performed first, the explicit dependence
of ${\cal A}_j$ on $(T,{\bf L})$ would have been obtained, i.e. a
gaussian spreading with $T$ in the direction ${\bf L}$. In section
\ref{section4.3} this spreading will be interpreted as the
longitudinal dispersion of the `wave packet' associated with the
propagating neutrino. We shall also see that computations
neglecting this dispersion lead to a result differing from
Eq.~(\ref{oscformula}) in the nonrelativistic limit (see
Eq.~(\ref{falsecohlength})). In contrast to the method of section
\ref{section4.3}, the question of dispersion does not even arise
in the derivation of Eqs.~(\ref{incohsumQFT})-(\ref{oscformula})
in which it is automatically taken into account.

Another feature of the oscillation formula (\ref{oscformula}) is
the increase of the coherence length when a long coherent
measurement in time is performed at the detector, even if the
neutrino `wave packets' have separated  spatially
\cite{kiers96,kiers98}. In that case, the energy uncertainty at
detection goes to zero, $\sigma_{e\SS D}\to0$, so that the
effective width also goes to zero, $\sigma_{peff}\to0$, and the
coherence length becomes infinite, $L^{coh}_{ij}\to\infty$. At
first sight, oscillations seem to vanish in that limit, because
the localization term depending on $\rho\sigma_{xeff}$ seems to
diverge \cite{giunti98}. If it were true, it would be impossible
to increase without limit the coherence length by performing long
coherent measurements. Note that this would be in contradiction
with stationary boundary condition models, which have a zero
energy uncertainty but an infinite coherence length. This apparent
contradiction can be cleared up by examining carefully the term
$\rho\sigma_{xeff}$ in the limit $\sigma_{e\SS D}\to0$. With the
help of Eqs.~(\ref{statlim}), (\ref{definitionrho}) and
(\ref{sigmaeff}), one sees that
$$
   |\rho| \, \sigma_{xeff}
   \stackrel{\sigma_{e\SS D} \to 0}{\longrightarrow}
   \frac{|v_{\SS D}|}{\sigma_{e\SS D}}
   \lesssim S_{\SS D}  \, ,
$$
where $S_{\SS D}$ was defined as the size of the macroscopic
detection region. Thus the localization term does not give a
stronger constraint than $S_{\SS D}\lesssim L^{osc}_{ij}$. This
constraint is always satisfied, as it is equivalent to the
constraint obtained by averaging the transition probability over
the production region. Therefore, the coherence length can be
increased without bound by more accurate energy measurements,
contrary to what was claimed in Refs.~\cite{giunti98,giunti98b}.
Note that this is not true if the accuracy of the 3-momentum
measurements is increased, as the localization makes the
oscillations vanish when the corresponding spatial uncertainty
becomes larger than the oscillation length. Of course the opposite
conclusions would be reached if experiments measured time, not
distance.

As a corollary, the stationary limit of the oscillation formula
(\ref{oscformula}) is well-defined. Stationary boundary
conditions, $v_{\SS P,D}=0$ and $\sigma_{e\SS P,D}=0$, lead to an
infinite effective width $\sigma_{xeff}$ and thus to an infinite
coherence length. However, the product $\rho \sigma_{xeff}$
remains finite, as we have just explained. For example, the
Grimus-Stockinger model \cite{grimus96} is obtained in the limit
$\rho=0$, $\sigma_{xeff}\!\to\!\infty$ with $\rho\sigma_{xeff}=0$.
The latter condition means that this model can be recovered from
the external wave packet model if the stationary limit has the
property $v_{\SS P,D}/\sigma_{e\SS P,D}=0$ (see discussion after
Eq.~(\ref{statlim})). Though condition (\ref{local1}) becomes
ineffective in that limit, the localization condition
(\ref{local2}) is still present. Thus there is no contradiction
between models with stationary boundary conditions and those with
external wave packets. The former type of model can be obtained
from the latter in a smooth limit.

In conclusion, if the observability conditions $|{\bf
p}_0\times{\bf l}|\lesssim\sigma_p$, $L\ll L^{coh}_{ij}$ and
$L^{osc}_{ij}\gg\sigma_x$ are satisfied, the oscillation formula
(\ref{oscformula}) reduces to the standard formula
(\ref{standard}), with the additional property of $1/L^2$
geometrical decrease.

\section{Other oscillation formulas}
\label{section4}

\subsection{The Jacob-Sachs theorem}

With result (\ref{oscformula}), we have been able to reconcile the
oscillation formulas proposed on the one hand by Giunti, Kim and
Lee \cite{giunti93,giunti98} and on the other hand by Grimus and
Stockinger \cite{grimus96}. More generally, the stationary
boundary conditions models have been shown to be special cases of
the external wave packet models. We have seen that
Eq.~(\ref{oscformula}) reduces to the standard oscillation formula
(\ref{standard}) if some observability conditions are verified;
otherwise oscillations vanish. However, other authors also working
in a quantum field framework have argued that formula
(\ref{oscformula}) is not valid for all experimental conditions.

First, Ioannisian and Pilaftsis \cite{ioannisian} derive a formula
for neutrino oscillations which exhibits a plane wave behaviour,
if the condition $L/\sigma_x \ll p_0/\sigma_p$ is satisfied (with
$\sigma_x\ll L^{osc}_{ij}$, as usual). By `plane wave
oscillations', these authors mean that the oscillation amplitude
${\cal A}_j$ does not decrease as the inverse of the distance,
that its phase depends on the direction ${\bf L}$ as ${\bf
p_j}\cdot {\bf L}$, and that the amplitude is not negligible in
directions other than along the neutrino momentum. If it were
true, the oscillation length measured by a specific detector would
depend on the direction of the total momentum of the initial
particles. To give a typical example, the above `plane wave
condition' is satisfied for atmospheric neutrinos if
$\sigma_x\gtrsim 10^{-2}\,$cm (with $L\!\sim\!1000\,$km and
$p\!\sim\!1\,$GeV). Unfortunately, this condition appears nowhere
in the treatment of the previous section. Actually, it is not
clear whether the Grimus-Stockinger theorem is valid under this
condition since its derivation uses the large $L$ limit.

Secondly, Shtanov \cite{shtanov} argues that the relative weights
of the oscillating exponentials can be strongly sensitive to the
neutrino masses, if the source and detector are strongly
localized. Since this author works in configuration space, the
derivation of the previous section cannot be useful to assess his
claim.

Thus, the examination of the claims made by Ioannisian, Pilaftsis
and Shtanov requires another method of evaluation of the amplitude
(\ref{ampli}). Instead of using the Grimus-Stockinger theorem to
integrate on the 3-momentum, we shall first perform the
integration on the energy with the Jacob-Sachs theorem
\cite{jacob}. This theorem is based on the assumption that the
energy spectrum of all incident particles is limited to a finite
range. Thus the overlap function $\psi(E,{\bf p})$ is distinct
from zero only for $p^2=E^2-{\bf p}^2$ within certain bounds (with
$E>0$). On this interval, $\psi(E,{\bf p})$ is taken to be
infinitely differentiable. In that case, the Jacob-Sachs theorem
says that the asymptotic value of the energy integral in
Eq.~(\ref{ampli}), when $T\to\infty$, is given by its residue at
the pole below the real axis. Thus the evaluation of the partial
amplitude (\ref{ampli}) with this theorem yields
\begin{equation}
   {\cal A}_j \cong \frac{\pi}{(2\pi)^4} \,  \int
   \frac{d^3p}{\sqrt{{\bf p}^2+m^2_j}} \;
   \psi(\sqrt{{\bf p}^2+m^2_j},{\bf p}) \;\,
   e^{ - i \phi_j({\bf p}) } \, ,
   \label{jacobsachsinteg}
\end{equation}
where $\phi_j({\bf p})=\sqrt{{\bf p}^2+m^2_j} \, T - {\bf p} \cdot
{\bf L}$. For gaussian external wave packets, the overlap function
is given by Eq.~(\ref{overlapPD}). In principle, this function
should be cut off outside the energy range determined by
experimental conditions so as to satisfy the conditions of the
Jacob-Sachs theorem. However these corrections are very small and
will be neglected in the computations.

It is interesting to note that the amplitude
(\ref{jacobsachsinteg}) is mathematically equivalent to the
amplitude obtained in the intermediate wave packet model
\cite{kayser}, in which the neutrino mass eigenstates are directly
represented by wave packets. The overlap function $\psi(\sqrt{{\bf
p}^2+m^2_j},{\bf p})$ corresponds to the wave function of the jth
mass eigenstate. Thus, it makes sense, in an external wave packet
model, to talk about mass eigenstate `wave packets' associated
with the propagating neutrino. Note however that the overlap
function takes into account not only the properties of the source,
but also of the detector, which is a bit unusual for a wave packet
interpretation.

\subsection{No-dispersion regime}
\label{section4.2}

\subsubsection{Choice of integration method}

The integration over the 3-momentum in Eq.~(\ref{jacobsachsinteg})
cannot be done analytically. Resorting to the explicit form
(\ref{overlapPD}) of the overlap function valid for gaussian
external wave packets, we see that the integral
(\ref{jacobsachsinteg}) can be approximated by means of an
asymptotic expansion for which two kinds of large parameters can
be used. On the one hand, $\sigma_{p\SS P,D}^{-2}$ and
$\sigma_{e\SS P,D}^{-2}$ are large parameters appearing in the
overlap function (\ref{overlapPD}). They suggest a second order
expansion of the integrand around the maximum ${\bf p}_j$ of the
overlap function, followed by a gaussian integration: this is
called {\it Laplace's method} \cite{erdelyi}. On the other hand,
$T$ and ${\bf L}$ are large parameters appearing in the phase.
They suggest a second order expansion of the integrand around the
stationary point ${\bf p}_{cl,j}$ of the phase, followed by a
gaussian integration: this is called the {\it method of stationary
phase} \cite{erdelyi}. The competition between these two
asymptotic behaviors implies a detailed study of the oscillation
of the phase around the average momentum ${\bf p}_j$. The
expansion of the phase in Eq.~(\ref{jacobsachsinteg}) should be
compared with the expansion of the overlap function. Although both
methods are expected to lead roughly to the same answer in the
case of the propagation of a single particle, it should be checked
whether the delicate compensation mechanism resulting in the
oscillation phase is independent of the method chosen.

The study of the overlap function amounts to the study of the
argument of the exponential, i.e. of the function $(f_{\SS
P}+f_{\SS D})({\bf p})$ (see Eq.~(\ref{overlapPD}) with
$p^0=\sqrt{{\bf p}^2+m^2_j}$). Using as in the previous section an
expansion in small mass differences $\epsilon=\delta m_j^2/E_0^2$,
the value ${\bf p}_j$ minimizing $f_{\SS P}+f_{\SS D}$ is given to
${\cal O}(\epsilon)$ by
\begin{equation}
  {\bf p}_j = {\bf p}_0 +
  \left(  \alpha {\bf u}_{\SS P} + \beta  {\bf u}_{\SS D} \right) \,
  \frac{\delta m_j^2}{2E_0} \, ,
  \label{momentumtilde}
\end{equation}
where
$$
  {\bf u}_{\SS P,D} =
  \frac{{\bf v}_0-{\bf v}_{\SS P,D}}{2\sigma_{e\SS P,D}}\, .
$$
The associated energy $E_j=\sqrt{{\bf p}_j^2+m_j^2}$ can be
expanded to ${\cal O}(\epsilon)$ as
\begin{equation}
  E_j = E_0 + \tilde\rho \, \frac{\delta m_j^2}{2E_0} \, ,
  \label{energytilde}
\end{equation}
where $\tilde\rho=1+\alpha{\bf v}_0\cdot{\bf u}_{\SS P}+\beta{\bf
v}_0\cdot{\bf u}_{\SS D}$. The values of the dimensionless
coefficients $\alpha$ and $\beta$ can be computed but we shall not
need their explicit expressions. It is sufficient to know that
$\tilde\rho\to0$ in the stationary limit. A velocity ${\bf
v}_j={\bf p}_j/E_j$ is also defined for future use.

We are now going to approximate the overlap function as a gaussian
and compute its three characteristic widths. At the extremum ${\bf
p}_j$, the Hessian matrix of $f_{\SS P}+f_{\SS D}$ reads to ${\cal
O}(\epsilon^0)$
\begin{eqnarray*}
  \Sigma^{ab} &\equiv&
  \frac{1}{2} \,
  \frac{\partial^2 (f_{\SS P}+f_{\SS D})}{\partial p^a\partial p^b}({\bf p}_j)
  \nonumber \\
  &=& \frac{\delta^{ab}}{4\sigma_p^2}
    + u_{\SS P}^a u_{\SS P}^b + u_{\SS D}^a u_{\SS D}^b \, ,
\end{eqnarray*}
where $\sigma_p$ is defined by Eq.~(\ref{definitionsigmap}). The
matrix $\Sigma^{ab}$ determines the range of ${\bf p}$ values for
which the overlap function $\psi(\sqrt{{\bf p}^2+m^2_j},{\bf p})$
is not negligible. As $\Sigma^{ab}$ is symmetric, it can be
diagonalized by an orthogonal coordinate transformation. The
eigenvalues of $\Sigma^{ab}$ are
\begin{eqnarray*}
  \sigma_x^2 &=& \frac{1}{4\sigma_p^2} \, ,
  \nonumber
  \\
  \sigma_{x\pm}^2 &=& \frac{1}{4\sigma_p^2}
  + \frac{1}{2} \left( {\bf u}_{\SS P}^2 + {\bf u}_{\SS D}^2 \right)
  \pm \frac{1}{2}
    \sqrt{\left( {\bf u}_{\SS P}^2 + {\bf u}_{\SS D}^2 \right)^2
          - 4 \left( {\bf u}_{\SS P} \times {\bf u}_{\SS D} \right)^2}
  \; .
\end{eqnarray*}
The eigenvector associated with $\sigma_x^2$ is in the direction
of ${\bf u}_{\SS P} \times {\bf u}_{\SS D}$, while the
eigenvectors associated with $\sigma_{x\pm}^2$ belong to the plane
defined by ${\bf u}_{\SS P}$ and ${\bf u}_{\SS D}$. In the limit
$|{\bf u}_{\SS P}|\gg|{\bf u}_{\SS D}|$ (resp. $|{\bf u}_{\SS
P}|\ll|{\bf u}_{\SS D}|$), the eigenvalues $\sigma_x^2$ and
$\sigma_{x-}^2$ become degenerate and the eigenvector associated
with $\sigma_{x+}^2$ becomes aligned with ${\bf u}_{\SS P}$ (resp.
${\bf u}_{\SS D}$). This is also the case in the limit of parallel
${\bf u}_{\SS P}$ and ${\bf u}_{\SS D}$. These limits are relevant
to the case of stationary boundary conditions which are examined
below.

Let us choose coordinate axes $({\bf e}_x,{\bf e}_y,{\bf e}_z)$
coinciding with the normalized eigenvectors associated with
$(\sigma_x^2,\sigma_{x-}^2,\sigma_{x+}^2)$ respectively. The
quantities $(\sigma_p^2,\sigma_{p-}^2,\sigma_{p+}^2)$  (with
$\sigma_{p\pm}\sigma_{x\pm}=1/2$) can be interpreted as the
momentum widths of the overlap function, since they give
constraints on the range of ${\bf p}$ values for which the overlap
function is non-negligible:
\begin{eqnarray}
  |p^x-p_j^x| &\lesssim& \sigma_p \;\; ,
  \nonumber \\
  |p^y-p_j^y| &\lesssim& \sigma_{p-} \; ,
  \nonumber \\
  |p^z-p_j^z| &\lesssim& \sigma_{p+} \; .
  \label{widthconstraint}
\end{eqnarray}

The case of the stationary limit is of special interest. Recall
that stationary boundary conditions are obtained in the external
wave packet model by taking ${\bf v}_{\SS P,D}\to0$ and
$\sigma_{e\SS P,D}\to0$ with $|{\bf v}_{\SS
P,D}|\lesssim\sigma_{e\SS P,D}S_{\SS P,D}$ (see
Eq.~(\ref{statlim})). In this limit, the axis ${\bf e}_z$ becomes
aligned with ${\bf v}_0$,
\begin{equation}
  v_0^{x,y} \sim v_{\SS P,D}^{x,y} \to 0  \, ,
  \label{statvo}
\end{equation}
and two eigenvalues become degenerate while the third diverges:
\begin{eqnarray}
  \sigma_{x-}^2 &\to& \sigma_x^2 \, ,
  \nonumber \\
  \sigma_{x+}^2 &\to&
  \frac{1}{4\sigma_p^2}
   + {\bf u}_{\SS P}^2 + {\bf u}_{\SS D}^2 \, ,
  \label{statwidth}
\end{eqnarray}
In other words, the transversal widths (i.e. in the directions
orthogonal to ${\bf p}_0$) are given in the stationary limit by
$\sigma_p$, while the longitudinal width (i.e. in the direction of
${\bf p}_0$) is given by $\sigma_{p+}\ll\sigma_p$. Note that the
asymptotic value of $\sigma_{x+}^2$ is very similar to the
definition (\ref{sigmaeff}) of $\sigma_{xeff}^2$.

The expansion of the overlap function has to be compared with the
expansion of the phase $\phi_j({\bf p})$ around ${\bf p}_j$, which
reads
\begin{equation}
  \phi_j({\bf p}) \cong \phi_j({\bf p}_j)
  + ({\bf v}_j T - {\bf L}) ({\bf p} - {\bf p}_j)
  + \frac{T}{2E_0} (p^a - p_j^a)R^{ab}(p^b - p_j^b)
  \, ,
  \label{expansionphase}
\end{equation}
where $R^{ab}=\delta^{ab}-v_0^av_0^b$. The second derivatives have
been evaluated to ${\cal O}(\epsilon^0)$.

Laplace's method will be preferred to the method of stationary
phase if the phase $\phi_j({\bf p})$ varies slowly over the `bump'
of the overlap function. As
$\sigma_p\geq\sigma_{p-}\geq\sigma_{p+}$, the variation of the
phase will become important in the first place along the direction
$p^x$, then in the direction $p^y$ and finally in the direction
$p^z$. The criterion for the use of Laplace's method in all three
directions $p^{x,y,z}$ will thus be determined by considering the
largest momentum width $\sigma_p$. The insertion of the
constraints (\ref{widthconstraint}) into the phase
(\ref{expansionphase}) yields first order conditions for a slowly
varying phase,
\begin{eqnarray}
  |v_j^x T - L^x| \sigma_p \; &\lesssim& 1 \, , \nonumber \\
  |v_j^y T - L^y| \sigma_{p-} &\lesssim& 1 \, , \nonumber \\
  |v_j^z T - L^z| \sigma_{p+} &\lesssim& 1 \, ,
  \label{cond1}
\end{eqnarray}
as well as a second order condition,
\begin{equation}
  \frac{T}{E_0} \, \sigma_p^2 \lesssim 1 \, ,
  \label{cond2}
\end{equation}
where we have used the property $q^aR^{ab}q^b\leq{\bf q}^2$ and
the fact that $\sigma_p$ is the largest width. For a given $T$
satisfying Eq.~(\ref{cond2}), it is always possible to find a
range of ${\bf L}$ values so that conditions (\ref{cond1}) are
satisfied. For other ${\bf L}$ values, the amplitude is negligible
as will be checked on the result (see Eq.~(\ref{Fdiag})). Thus the
criterion allowing to choose the integration method is determined
by Eq.~(\ref{cond2}): the integration on ${\bf p}$ will be done by
Laplace's method if $T\lesssim E_0/\sigma_p^2$ or, equivalently
(with $L=|{\bf L}|$ and $p_0=|{\bf p}_0|$), if
\begin{equation}
   L\lesssim \frac{p_0}{\sigma_p^2} \, ,
  \label{ipcondition}
\end{equation}
since conditions (\ref{cond1}) impose the relation ${\bf L}\cong
{\bf v}_0T$ as long as $\sigma_{x+}\ll L$.

For $\sigma_{x+}\gtrsim L$ (stationary limit), we now show that
condition (\ref{ipcondition}) is directly obtained without going
through condition (\ref{cond2}). The overlap function imposes in
that limit that $|{\bf p}|=\sqrt{E_0^2-m_j^2}$, so that we are
left with an angular integration with the angular part of the
integrand given by
\begin{equation}
  \exp\left( \frac{{\bf p}\cdot{\bf p}_0}{2\sigma_p^2}
            + i {\bf p}\cdot{\bf L}
      \right) \, .
  \label{angular}
\end{equation}
Condition (\ref{ipcondition}) shows that the angular variation of
the phase in (\ref{angular}) is slow with respect to the angular
variation of the overlap function, in which case Laplace's method
will give good results. Therefore condition (\ref{ipcondition})
constitutes a good criterion for the use of Laplace's method
whether the stationary limit is taken or not.

Condition (\ref{ipcondition}) is usually not verified in
oscillation experiments, because $L/\sigma_x\gg p_0/\sigma_p$ in
most cases. This condition is the same than the one under which
Ioannisian and Pilaftsis \cite{ioannisian} obtain `plane wave'
oscillations.

\subsubsection{Amplitude}

If condition (\ref{ipcondition}) is satisfied, the evaluation of
the integral (\ref{jacobsachsinteg}) can be done by Laplace's
method and yields
\begin{equation}
  {\cal A}_j \sim \sigma_p\sigma_{p-}\sigma_{p+} \,
  \exp \left(
      - i E_j T + i {\bf p}_j \cdot {\bf L}
      - \left( \frac{\delta m_j^2}{4\tilde\sigma_mE_0} \right)^2
      - F_j(T) \right) \, ,
  \label{nodispersion}
\end{equation}
where the third term in the exponential comes from $(f_{\SS
P}+f_{\SS D})({\bf p}_j)$; the parameter $\tilde\sigma_m$ can be
computed and has the dimension of a momentum width. The function
$F_j(T)$ is defined by
\begin{equation}
  F_j(T)= \frac{1}{4}
          ({\bf v}_j T - {\bf L})^t
          \left( \Sigma + i\frac{T}{2E_0} R \right)^{-1}
          ({\bf v}_j T - {\bf L}) \, ,
  \label{definF}
\end{equation}
where $\Sigma^{ab}$ and $R^{ab}$ are considered as matrices. In
the framework of the wave packet interpretation developed after
Eq.~(\ref{jacobsachsinteg}), the function $\exp(- F_j(T))$ plays
the part of the space-time envelope of the wave packet associated
with the jth neutrino mass eigenstate. The elements of the matrix
$Re(\Sigma+i\frac{T}{2E_0}R)^{-1}$ constrain the extent of the
wave packet envelope in space-time. As $T$ increases, the wave
packet spreads because of the $i\frac{T}{E_0}R$ term. Thus the
dispersion of the wave packet is due to the second order term in
the expansion of the phase $\phi_j({\bf p}_j)$. Therefore,
condition (\ref{cond2}) or, equivalently, condition
(\ref{ipcondition}) means that dispersion has not yet begun in any
direction, transversal or longitudinal. For that reason, the range
of ${\bf L}$ values defined by $L\lesssim p_0/\sigma_p^2$ will be
called the {\it no-dispersion regime}. Of course, this
interpretation is not valid for $\sigma_{x+}\gtrsim L$, in which
case the propagation time $T$ becomes indeterminate and dispersion
loses its meaning.

Now that the origin of dispersion has been clarified, the term in
Eq.~(\ref{definF}) including $R$ can be neglected with respect to
$\Sigma$. Moreover, we approximate ${\bf v}_j$ by ${\bf v}_0$ in
$F_j(T)$. In comparison with the oscillation formula
(\ref{oscformula}) derived in section \ref{section3}, this
approximation will lead to the absence of the coherence-length
term, since this term exclusively arises, when the dispersion is
neglected, from the velocity difference ${\bf v}_i-{\bf v}_j$.
Dropping the index $j$, the wave packet envelope in
Eq.~(\ref{nodispersion}) can then be written in the coordinate
system diagonalizing $\Sigma$ as
\begin{equation}
  F(T)=    \frac{(v_0^x T - L^x)^2}{4\sigma_x^2}
         + \frac{(v_0^y T - L^y)^2}{4\sigma_{x-}^2}
         + \frac{(v_0^z T - L^z)^2}{4\sigma_{x+}^2} \, ,
  \label{Fdiag}
\end{equation}
which shows that the conditions (\ref{cond1}) assumed for
Laplace's method are required to obtain a non-negligible amplitude
${\cal A}_j$. If $\sigma_{x+}\ll L$, condition (\ref{Fdiag})
imposes the relation
$$
  {\bf v}_0T = {\bf L} + {\cal O}(\sigma_{x+}) \, .
$$
In that case, the phase of the amplitude (\ref{nodispersion}) is
given to ${\cal O}(\epsilon)$ by
\begin{equation}
  E_j T - {\bf p}_j \cdot {\bf L} =
  E_0 T - {\bf p}_0 \cdot {\bf L}
  + \frac{\delta m_j^2}{2p_0} \,
  \left( L + {\cal O}(\sigma _{x+}) \right) \, ,
  \label{phasenodisp}
\end{equation}
where the expansions (\ref{momentumtilde}) and (\ref{energytilde})
have been used. This result shows that the phase difference
between the amplitude ${\cal A}_i$ and ${\cal A}_j^*$ is equal to
the standard oscillation phase present in Eq.~(\ref{standard}).

What is the form of the amplitude (\ref{nodispersion}) in the case
of stationary boundary conditions ($\sigma_{x+}\gtrsim L$)? This
question is important since such conditions were assumed by
Ioannisian and Pilaftsis in their derivation of `plane wave
oscillations' \cite{ioannisian}. Let us first examine the wave
packet envelope $F(T)$. In the stationary limit (see
Eq.~(\ref{statwidth})), $\sigma_{x+}$ becomes larger than $L^z$ so
that the $z$ component in Eq.~(\ref{Fdiag}) imposes that $|v_0^z|
T\lesssim\sigma_{x+}$. Putting this condition together with
Eqs.~(\ref{statlim}) and (\ref{statvo}), we obtain
$$
  |v_0^{x,y}|T
  \lesssim \frac{|v_{\SS P,D}^{x,y}|}{|v_0^z|} \, \sigma_{x+}
  \lesssim S_{\SS P,D} \, .
$$
Thus the wave packet envelope (\ref{Fdiag}) gives the following
constraint in the stationary limit:
\begin{equation}
  |L^{x,y}| \lesssim S_{\SS P,D} \, ,
  \label{Fstat}
\end{equation}
which means that, for propagation distances much larger than the
size of the source and detector ($L\gg S_{\SS P,D}$), the
amplitude (\ref{nodispersion}) is negligible unless the direction
of observation ${\bf L}$ is nearly parallel to the average
neutrino momentum ${\bf p}_0$. Thus this constraint is valid
whether the stationary limit is taken or not.

Let us now examine the phase of the amplitude (\ref{nodispersion})
in the stationary limit. The expansions (\ref{momentumtilde}) and
(\ref{energytilde}) and the property $\tilde\rho\to0$ show that
\begin{eqnarray*}
  E_j &\to& E_0 \, , \\
  {\bf v}_0 \cdot {\bf p}_j &\to&
  {\bf v}_0 \cdot {\bf p}_0 - \frac{\delta m_j^2}{2E_0} \, .
\end{eqnarray*}
Using ${\bf L} = {\bf v}_0L/|{\bf v}_0|$ (with correction terms
given by Eq.~(\ref{Fstat})), the phase of the amplitude is given
to ${\cal O}(\epsilon)$ by
\begin{equation}
  E_j T - {\bf p}_j \cdot {\bf L} =
  E_0 T - {\bf p}_0 \cdot {\bf L}
  + \frac{\delta m_j^2}{2p_0} \,
  \left(L + {\cal O}(S_{\SS P,D})\right) \, ,
  \label{phasenodispstat}
\end{equation}
which leads again to the standard oscillation phase present in
Eq.~(\ref{standard}) since $L\gg S_{\SS P,D}$. Eqs.~(\ref{Fstat})
and (\ref{phasenodispstat}) also show that apart from a factor
$exp(-iE_0 T)$ which can be factorized from the sum over the mass
eigenstates, the amplitude (\ref{nodispersion}) is independent of
time in the stationary limit.

We have thus demonstrated that the amplitude (\ref{nodispersion}),
derived under condition (\ref{ipcondition}), is negligible in
directions other than the neutrino average momentum whether the
stationary limit is taken or not. Therefore the oscillation phase
has the standard form given in Eq.~(\ref{standard}) and no `plane
wave oscillations' can be observed, contrary to the claim made by
Ioannisian and Pilaftsis \cite{ioannisian}. Also, the absence of
the $1/T$ (or $1/L$) factor in Eq.~(\ref{nodispersion}), as noted
by the same authors, is easily understood by noting that the
absence of dispersion entails that the amplitude does not decrease
with the distance.

\subsubsection{Probability}

As in section \ref{section3}, the last step towards the
oscillation formula consists in computing the time average of the
transition probability, which is a gaussian integral on $T$:
\begin{equation}
  \int dT \; {\cal A}_i{\cal A}_j^* \sim
  \exp \left(
      - i\phi_{ij}(T_0)
      - \frac{(\delta m_i^2)^2+(\delta m_j^2)^2}{16\tilde\sigma_m^2E_0^2}
      - \frac{(E_i-E_j)^2}{4F''(T_0)} - 2F(T_0) \right) \, ,
  \label{averagenodisp}
\end{equation}
where $T_0$ is the solution of $F'(T_0)=0$, while the phase
$\phi_{ij}(T_0)$ is given by
\begin{equation}
  \phi_{ij}(T_0) = (E_i-E_j) T_0 - ({\bf p}_i-{\bf p}_j) \cdot {\bf L}
  \, , \label{phaseprobanodisp}
\end{equation}
The function $F(T)$ can be rewritten as
\begin{equation}
  F(T)= \frac{(\tilde{\bf v}_0 T - \tilde{\bf L})^2}{4\sigma_x^2}
  \, ,
\end{equation}
where
\begin{eqnarray*}
  \tilde{\bf v}_0 &=& \sigma_x\sqrt{\Sigma^{-1}} \, {\bf v}_0 \, ,
  \\
  \tilde{\bf L} &=& \sigma_x\sqrt{\Sigma^{-1}} \, {\bf L} \, ,
\end{eqnarray*}
with
$\Sigma^{-1}=diag(\sigma_x^{-2},\sigma_{x-}^{-2},\sigma_{x+}^{-2})$.
It is then easy to compute
\begin{eqnarray}
  T_0 &=& \frac{\tilde{\bf v}_0\cdot\tilde{\bf L}}{\tilde{\bf v}_0^2}
  \, , \label{valueTO} \\
  F(T_0) &=&
  \frac{(\tilde{\bf v}_0\times\tilde{\bf L})^2}{4\sigma_x^2\,\tilde{\bf v}_0^2}  \, ,
  \label{directioncond} \\
  F''(T_0) &=& \frac{\tilde{\bf v}_0^2}{2\sigma_x^2} \, .
  \label{secderiv}
\end{eqnarray}
We now show that results
(\ref{phasenodisp})-(\ref{phasenodispstat}), obtained at the level
of the amplitude, can be reproduced by a careful analysis of
Eq.~(\ref{averagenodisp}). This is not completely trivial as the
integration variable $T$ becomes indeterminate in the stationary
limit.

The insertion of expression (\ref{directioncond}) in
Eq.~(\ref{averagenodisp}) shows that $F(T_0)$ plays the role of a
directional constraint. More precisely, $\int dT{\cal A}_i{\cal
A}_j^*$ is non-negligible if $\tilde{\bf L}$ is nearly parallel to
$\tilde{\bf v}_0$ or, for $\sigma_{x+}\ll L$, if ${\bf L}$ is
nearly parallel to ${\bf v}_0$:
\begin{equation}
  {\bf L} = \frac{{\bf v}_0}{|{\bf v}_0|} \, L
  + {\cal O}(\sigma_{x+})  \, .
  \label{substi}
\end{equation}
With the substitution (\ref{substi}), the value of $T_0$ given by
Eq.~(\ref{valueTO}) becomes $T_0=L/|{\bf v}_0|+{\cal
O}(\sigma_{x+})$, so that the phase (\ref{phaseprobanodisp})
becomes
\begin{equation}
  \phi_{ij}(T_0) = \frac{\delta m_{ij}^2}{2p_0} (L + {\cal O}(\sigma_{x+})) \, ,
  \label{phaseT0}
\end{equation}
which is equal to the standard oscillation phase present in
Eq.~(\ref{standard}) and equivalent to result (\ref{phasenodisp}).

If $\sigma_{x+}\gtrsim L$, $\tilde{\bf v}_0$ and $\tilde L^z$ tend
to zero so that $F(T_0)$ should directly be studied as a function
of ${\bf v}_0$ and ${\bf L}$. General conclusions for arbitrary
$\sigma_{x\pm}$ can be drawn from the study of the quadratic form
in $(L^x,L^y,L^z)$ associated with $F(T_0)=1$. This analysis shows
that there is an eigenvalue $s_3=0$ corresponding to an
eigenvector along ${\bf v}_0$. The two other eigenvalues $s_{1,2}$
are positive (with $s_2\leq s_1$) , so that the surface $F(T_0)=1$
in $(L^x,L^y,L^z)$-space is a cylinder of elliptical section with
an axis along ${\bf v}_0$. This geometrical picture can be
interpreted as imposing that the components of ${\bf L}$
orthogonal to ${\bf v}_0$ should be smaller than $\sqrt{1/s_2}$,
whereas there is no constraint at all on the component of ${\bf
L}$ along ${\bf v}_0$. In the stationary limit, the lengthy
expressions of the non-zero eigenvalues become
\begin{eqnarray}
  s_1 &\to& \frac{1}{4\sigma_x^2} \, ,
  \nonumber \\
  s_2 &\to& \frac{1}{4\sigma_x^2} \,
  \frac{\sigma_x^2\,(v_0^z)^2}
       {\sigma_{x+}^2(v_0^x)^2+\sigma_{x+}^2(v_0^y)^2+\sigma_{x}^2(v_0^z)^2} \, .
  \label{s12stat}
\end{eqnarray}
The properties of the stationary limit, namely
Eqs.~(\ref{statlim}), (\ref{statvo}) and (\ref{statwidth}), lead
to the bound $\sqrt{1/s_2}\lesssim S_{\SS P,D}$. In the stationary
limit, the components $L^{x,y}$ (which are then orthogonal to
${\bf v}_0$) should thus be smaller than $S_{\SS P,D}$. Thus the
time-averaged probability is always negligible in directions other
than along the average neutrino momentum ${\bf p}_0$. Note that
this condition reproduces the constraint (\ref{Fstat}) derived in
the study of the stationary limit of the amplitude
(\ref{nodispersion}). The phase $\phi_{ij}(T_0)$ can then be
evaluated as in Eq.~(\ref{phasenodispstat}), leading to the
oscillation phase
\begin{equation}
  \phi_{ij}(T_0) = \frac{\delta m_{ij}^2}{2p_0} (L + {\cal O}(S_{\SS P,D})) \, ,
  \label{phaseT0stat}
\end{equation}
which is again equal to the standard oscillation phase present in
Eq.~(\ref{standard}) and equivalent to result
(\ref{phasenodispstat}).

Putting together the results (\ref{averagenodisp}),
(\ref{directioncond}), (\ref{secderiv}) and (\ref{phaseT0}) (or
(\ref{phaseT0stat})), the flavor-mixing transition probability
(\ref{proba}) for a propagation distance satisfying condition
(\ref{ipcondition}) can be written as
\begin{eqnarray}
  && {\cal P}_{\alpha \to \beta}({\bf L}) \sim
  \exp \left( - \frac{(\tilde{\bf v}_0\times\tilde{\bf L})^2}
                     {2\sigma_x^2\,\tilde{\bf v}_0^2}
       \right)
  \sum_{i,j} V_{i\alpha} \, V_{i\beta}^* \, V_{j\alpha}^* \, V_{j\beta}
  \nonumber \\
  && \hspace{2cm} \times
  \exp \left(
      - 2\pi i\frac{L}{L_{ij}^{osc}}
      - 2\pi^2\left( \frac{\tilde\rho\tilde\sigma_{xeff}}{L_{ij}^{osc}} \right)^2
      - \frac{(\delta m_i^2)^2+(\delta m_j^2)^2}{16\tilde\sigma_m^2E_0^2}
      \right) \, ,
  \label{probanodisp}
\end{eqnarray}
where $\tilde\sigma_{xeff}=\sigma_x|{\bf v}_0|/|\tilde{\bf v}_0|$
and $L_{ij}^{osc}$ is given by Eq.~(\ref{osclength}).

This oscillation formula is very similar to
Eq.~(\ref{oscformula}). There are three main differences, two of
which have already been mentioned. First there is no
coherence-length term in Eq.~(\ref{probanodisp}), which is due to
the neglect of the terms beyond ${\cal O}(\epsilon^0)$ in the
evaluation of $F(T_0)$ (see Eq.~(\ref{Fdiag})). Secondly, the
geometrical decrease in $1/L^2$ is lacking in
Eq.~(\ref{probanodisp}), which is explained by the fact that the
dispersion of the neutrino `wave packet' is not yet significant.
Finally, the directional constraint present in
Eq.~(\ref{oscformula}) restricts the neutrino propagation to a
cone of angle $arcsin(\sigma_p/p_0)$, whereas the directional
constraint in Eq.~(\ref{probanodisp}) confines the propagation to
a cylinder of radius $r$ (with $\sigma_x\lesssim r\lesssim S_{\SS
P,D }$). This different behavior is also a result of the absence
of dispersion for $L\lesssim p_0/\sigma_p^2$.

Before closing this section, it is interesting to understand why a
directional constraint is missing in Ref.~\cite{ioannisian} for
$L\lesssim p_0/\sigma_p^2$, as this fact explains the `plane wave
oscillations' result. At the end of their computations, the
authors of Ref.~\cite{ioannisian} obtain an amplitude ${\cal A}_j$
whose dominant term depends on $exp(ip_j|{\bf L'}|)$, where $|{\bf
L'}|=\sqrt{{\bf L'}^2}$ is the complex `norm' of a complex vector
${\bf L'}={\bf L}-2i\sigma_x^2{\bf p}_0$ (we have translated their
results in our notation through the correspondence $q_j\to p_j$,
$\vec k\to{\bf p}_0$, $\delta l^2\to4\sigma_x^2$, $\vec L\to{\bf
L'}$, $\vec l\to{\bf L}$). For $L\ll p_0/\sigma_p^2$, the quantity
$|{\bf L'}|$ can be expanded and the argument of the exponential
reads to second order
$$
  ip_j|{\bf L}-2i\sigma_x^2{\bf p}_0| \cong
  2\sigma_x^2p_0p_j
  + i p_j\frac{{\bf p}_0\cdot{\bf L}}{p_0}
  -\frac{p_j}{4\sigma_x^2p_0}\,
  \left( L^2 - \frac{({\bf p}_0\cdot{\bf L})^2}{p_0^2} \right)
  \, .
$$
The two last terms were neglected in Ref.~\cite{ioannisian} though
$L\gg\sigma_x$. They lead to the following directional constraint:
\begin{equation}
  \exp\left( - \frac{({\bf v}_0\times{\bf L})^2}{4\sigma_x^2v_0^2} \right)
  \, , \label{ipdirconstraint}
\end{equation}
where the factor 4 becomes a factor 2 when the amplitude is
squared. Therefore the result of Ref.~\cite{ioannisian} also
includes for $L\ll p_0/\sigma_p^2$ a directional constraint
forbidding plane wave oscillations.

Note that the stationary limit assumption $|{\bf v}_{\SS
P,D}|/\sigma_{e\SS P,D}=0$ leads to $s_1=s_2$ (see
Eq.~(\ref{s12stat})), so that the constraint
(\ref{ipdirconstraint}) becomes equal to the one present in
Eq.~(\ref{probanodisp}). Thus the condition $|{\bf v}_{\SS
P,D}|/\sigma_{e\SS P,D}=0$ seems generic for models with
stationary boundary conditions, since it was also applied in the
case of the Grimus-Stockinger model (see discussion after
Eq.~(\ref{statlim}) and at the end of section \ref{section3}).

In conclusion, the analysis of the amplitude (\ref{nodispersion})
(as well as of the transition probability (\ref{probanodisp}))
derived in the external wave packet model under condition
(\ref{ipcondition}) shows that `plane wave oscillations' do not
exist. This negative result is confirmed by a reexamination of the
formula derived in Ref.~\cite{ioannisian}. Besides providing a
rigorous method to compute the transition probability
(\ref{probanodisp}), the external wave packet model has the
advantage of associating a clear physical picture to the different
stages of computations. For example, condition (\ref{ipcondition})
can be interpreted as meaning that the wave packet associated with
the neutrino has not yet begun to spread in any direction.
Finally, the oscillation formula (\ref{probanodisp}) reduces to
the standard formula (\ref{standard}) if the observability
conditions $|{\bf v}_0\times{\bf L}|\lesssim v_0\sigma_x$ and
$L^{osc}_{ij}\gg\sigma_x$ are satisfied.

\subsection{Transversal- and longitudinal-dispersion regimes}
\label{section4.3}

Let us now assume that condition (\ref{ipcondition}) is not
satisfied, i.e.\ $L\gtrsim p_0/\sigma_p^2$. In that case,
Laplace's method cannot be used to integrate on all three
components $p^{x,y,z}$ in the amplitude (\ref{jacobsachsinteg})
since dispersion becomes significant. However the spreading of the
neutrino `wave packet' is not identical in all directions. More
specifically, the onset of dispersion in the direction ${\bf p}_0$
can be delayed by two factors. First, the matrix element $R^{ab}$
present in Eq.~(\ref{expansionphase}) leads to a relativistic
contraction (of $1-{\bf v}_0^2$) in the direction ${\bf p}_0$ of
the dispersion of the neutrino `wave packet' (see
Eq.~(\ref{definF})). Secondly, the momentum width along ${\bf
p}_0$ is given for $\sigma_{e\SS P,D}\ll\sigma_x$ (i.e. in the
stationary limit (\ref{statwidth})) by a vanishing $\sigma_{p+}$.
Thus Laplace's method will be valid for a longer time $T$ in the
direction ${\bf p}_0$ than in directions transverse to this
vector. For this reason, the choice of the integration method in
the direction ${\bf p}_0$ will be postponed for a short while,
while the method of stationary phase will be preferred  for
momentum integrations in directions transverse to ${\bf p}_0$.

Let the $z$ axis be along ${\bf L}$, i.e.\ ${\bf L}=L\,{\bf e}_z$.
As in section \ref{section4.2}, the examination of the amplitude
(\ref{jacobsachsinteg}) shows that the quick variation of the
phase averages the amplitude to zero unless ${\bf p}_0$ and ${\bf
L}$ are nearly parallel. The method of stationary phase will thus
be applied in directions $p^{x,y}$, the stationary points of which
are given by $p^x=p^y=0$.  The result of the method of stationary
phase for the transverse momenta in (\ref{jacobsachsinteg}) can be
written as follows:
\begin{equation}
  {\cal A}_j \sim \frac{g({\bf l})}{T-i\mu} \,
  \int dp \; e^{ - i \phi_j(p) - f_j(p) } \, ,
  \label{longitudinalamplitude}
\end{equation}
where $p\equiv p^z$ and $\mu=E_0/2\sigma_p^2$. The geometrical
constraint $g({\bf l})$ is given by Eq.~(\ref{geom}), while the
momentum width $\sigma_p$ is defined by
Eq.~(\ref{definitionsigmap}). The phase $\phi_j(p)$ is given by
\begin{equation}
  \phi_j(p)= \sqrt{p^2+m_j^2} \, T - pL \, .
  \label{phasej}
\end{equation}
The original overlap function is partly included in $g({\bf l})$
and partly in
\begin{equation}
  f_j(p)= f_{j\SS P}(p) + f_{j\SS D}(p) \, ,
  \label{definitionfj}
\end{equation}
with
$$
  f_{j\SS P}(p) =
  \frac{ (p-p_0)^2}{4\sigma_{p \SS P}^2}
  +\frac{ \left( \sqrt{p^2+m_j^2} - E_0 -( p-p_0) v_{\SS P} \right)^2}
         { 4\sigma_{e\SS P}^2 } \, ,
$$
where $v_{\SS P}= v_{\SS P}^z$ and $p_0=p_0^z$. The energy $E_0$
has been redefined so as to absorb a factor $p_0^x v_{\SS P}^x +
p_0^y v_{\SS P}^y$. The definition of $f_{j\SS D}(p)$ is similar.
As $T\gtrsim E_0/\sigma_{peff}^2$, the term $1/(T-i\mu)$ can be
approximated by $1/T$ so as to provide a prefactor $1/L^2$ in the
transition probability. This expected geometrical decrease is seen
to originate in the transverse dispersion of the wave packet
corresponding to the oscillating particle. For future use, we
define a reference mass $m_0$ by $m_0^2=E_0^2-p_0^2$ and a
velocity $v_0$ by $v_0=p_0/E_0$.

As regards the longitudinal momentum integral, it is tempting to
proceed as in section \ref{section3}, i.e.\ to compute first the
time average of the transition probability before integrating on
$p$. However, the prefactor $1/(T-i\mu)$ present in the amplitude
(\ref{longitudinalamplitude}) gives in that case a delta function
look-alike of width $\mu^{-1}\sim\sigma_p^2/E_0$, introducing an
additional momentum uncertainty which is larger than the mass
difference $\delta m^2/E$ since $L^{osc}_{ij}\gtrsim
p_0/\sigma_p^2$. For this reason, it is preferable to avoid this
shortcut (note that it yields the same final answer as given by
the following method). Moreover, it is interesting for the
physical interpretation to postpone the time average, so as to
obtain the explicit dependence of the amplitude on time and
distance. As in the case of transverse momenta integrals, the
choice of the method to perform the longitudinal momentum
integration (\ref{longitudinalamplitude}) is done by comparing the
expansions of the phase and of the overlap function around the
value $p_j$ for which $f_j(p)$ is extremal. One obtains to ${\cal
O}(\epsilon)$
$$
  p_j=p_0 + (\rho-1)\, \frac{\delta m_j^2}{2p_0} \, ,
$$
where the dimensionless number $\rho$ is defined by
Eq.~(\ref{definitionrho}). The associated energy
$E_j=\sqrt{p_j^2+m_j^2}$ and velocity $v_j=p_j/E_j$ are given to
${\cal O}(\epsilon)$ by
\begin{eqnarray}
  E_j &=& E_0 + \rho \frac{\delta m_j^2}{2E_0} \, ,
  \nonumber
  \\
  v_j &=& v_0
  + \left(\rho(1-v_0^2)-1 \right)\frac{\delta m_j^2}{2p_0E_0} \, .
  \label{expansionvj}
\end{eqnarray}
The expansions of $f_j(p)$ and $\phi_j(p)$ are given by
\begin{eqnarray}
  f_j(p) &\cong& \frac{(p-p_j)^2}{4\sigma_{peff}^2} \, ,
  \nonumber
  \\
  \phi_j(p)&\cong& \phi_j(p_j) + (v_jT-L)(p-p_j)
  + \frac{m_j^2T}{2E_0^3} (p-p_j)^2 \, ,
  \label{expansionphij}
\end{eqnarray}
where $\sigma_{peff}$ is given by Eq.~(\ref{sigmaeff}) and can
again be interpreted as the longitudinal width of the overlap
function. Laplace's method will be used if the phase
(\ref{expansionphij}) varies slowly over the width
$\sigma_{peff}$, i.e. if the two following conditions are
satisfied
\begin{eqnarray}
  |v_jT-L|2\sigma_{peff} &\lesssim& 1 \, ,
  \label{constraint1}
  \\
  \frac{m_j^2T}{2E_0^3} 4\sigma_{peff}^2 &\lesssim& 1 \, .
  \label{constraint2}
\end{eqnarray}
As in section \ref{section4.2}, the first order constraint
(\ref{constraint1}) is included in the result of Laplace's method.
Thus the criterion allowing to choose between Laplace and
stationary phase methods is given by Eq.~(\ref{constraint2}). In
other words, it is better to use Laplace's method if $T$ is
smaller than a {\it dispersion time} $T^{disp}_j$ defined by
\begin{equation}
  T^{disp}_j = \frac{E_0^3}{2m_j^2\sigma_{peff}^2} \, .
  \label{dispersiontime}
\end{equation}
The term `dispersion time' is justified by the fact that it is the
time at which the longitudinal dispersion of the amplitude becomes
important, more precisely twice the initial size. A {\it
dispersion length} $L^{disp}_j$ can be defined by $L^{disp}_j=v_0
\, T^{disp}_j$. The distance range $p_0/\sigma_{peff}^2\lesssim
L\lesssim L^{disp}_j$ will be called the {\it tranverse-dispersion
regime}. For $L\gtrsim L^{disp}_j$, the stationary phase method is
more accurate: this distance range will be called the {\it
longitudinal-dispersion regime}. Various estimates of the
dispersion length are given in \cite{beuthe}, showing that the
concept of dispersion length is relevant to nonrelativistic
particles as well as to supernova neutrinos, and possibly to solar
neutrinos.

In the transverse-dispersion regime, the evaluation of the
amplitude (\ref{longitudinalamplitude}) as a gaussian integral
around $p_j$ gives
\begin{equation}
  {\cal A}_j \sim
  \frac{g({\bf l}) \sigma_{peff}}{ T\sqrt{1+i T/T^{disp}_j} } \,
  \exp \left(
  -i E_j T + i p_j L
  - \left( \frac{\delta m_j^2}{4\sigma_m E_0} \right)^2
  - \frac{1}{1+iT/T^{disp}_j} \frac{(v_jT-L)^2}{4\sigma_{xeff}^2}
  \right) \, .
  \label{transdisp}
\end{equation}
The width $\sigma_m$ is defined by Eq.~(\ref{sigmam}). The
amplitude (\ref{transdisp}) behaves as a wave packet of group
velocity $v_j$ and space-time extent
$(1+(T/T_j^{disp})^2)^\frac{1}{2}\,\sigma_{xeff}$. If the
longitudinal dispersion is neglected ($T_j^{disp}=\infty$), the
amplitude (\ref{transdisp}) is similar to Eq.~(18) of
Ref.~\cite{giunti98}.

The following step consists in computing the time averaged
transition probability (\ref{proba}). If dispersion is taken into
account ($T_j^{disp}<\infty$), a tedious gaussian integration by
Laplace's method (see \cite{beuthe}) yields the oscillation
formula (\ref{oscformula}). If the dispersion is neglected
($T_j^{disp}=\infty$), a much shorter computation leads to the
same oscillation formula (\ref{oscformula}), except that the
following substitution has to be made:
\begin{equation}
   L^{coh}_{ij} \to
   \frac{L^{coh}_{ij}}{|\rho(1-v_0^2)-1 |}
    \;\;\; (FALSE) \, .
   \label{falsecohlength}
\end{equation}
The incorrect multiplying factor has its origin in
Eq.~(\ref{expansionvj}): in the limit $T_j^{disp}=\infty$, the
coherence length term arises only from the difference between the
group velocities $v_i$ and $v_j$. However the factor
$|\rho(1-v_0^2)-1 |$ tends to 1 in the relativistic limit, so that
the substitution (\ref{falsecohlength}) becomes trivial. This
observation explains why our result (\ref{oscformula}) coincides
with Eq.~(26) of Ref.~\cite{giunti98}, as the authors of this
article, while neglecting dispersion, consider only relativistic
neutrinos. Note however that even relativistic neutrinos spread at
large distances so that a calculation neglecting dispersion such
as in Ref.~\cite{giunti98} is only valid for $L\lesssim
L_j^{disp}$.

At sufficiently large distance, dispersion becomes significant and
all neutrinos propagating freely enter into the
longitudinal-dispersion regime. In this regime, we have argued
that the integral (\ref{longitudinalamplitude}) should be
evaluated with the method of stationary phase. The stationary
point of the phase $\phi_j(p)$ is given by
$$
  p_{cl,j} = m_j \frac{v_{cl}}{\sqrt{1-v_{cl}^2}} \, ,
$$
where $v_{cl}=L/T$. It can be interpreted as the classical
momentum of a particle of mass $m_j$, travelling at the classical
velocity $v_{cl}$. The evaluation of the amplitude
(\ref{longitudinalamplitude}) as a gaussian integral around
$p_{cl,j}$ gives
\begin{equation}
  {\cal A}_j \sim
  \frac{g({\bf l})\sigma_{peff}}{ T\sqrt{1+i T/T^{disp}_j} } \,
  \exp \left( - i m_j \, \sqrt{T^2-L^2} - f_j(p_{cl,j})
              + \sigma_{peff}^2 \,
                \frac{ \left( f_j'(p_{cl,j}) \right)^2 }{1+i T/T^{disp}_j}
       \right)  \, ,
  \label{longdisp}
\end{equation}
where $f_j(p)$ is defined by Eq.~(\ref{definitionfj}), $f_j'(p)$
refers to its derivative. The wave packet interpretation of the
amplitude (\ref{longdisp}) is not obvious but the shape of the
associated wave packet can be studied by an expansion around the
maximum of the amplitude. Once more, the time average of the
transition probability (\ref{proba}) is computed by Laplace's
method (see Ref.~\cite{beuthe}), yielding again the oscillation
formula (\ref{oscformula}).

It is striking that the two different methods of approximation
used in this section, a priori valid in different regimes, give
the same oscillation formula, which also coincides with the result
obtained with the approximation method used in section
\ref{section3}, Eq.~(\ref{oscformula}). The dispersion length does
not play any special role in the final result. Thus, each method
is accurate enough to be extended to the whole range of distances.
However, the physical interpretation of the terms appearing in the
oscillation formula (\ref{oscformula}) depends on the relative
values of $L$ and $L^{disp}_{i,j}$. From the detailed analysis
done in \cite{beuthe}, these terms can be interpreted as follows:
\begin{itemize}
\item
In the transverse-dispersion regime, the coherence-length
term mainly arises from the progressive spatial separation of
`wave packets', due to their different group velocities. The
localization term mainly arises from the initial spread of the
`wave packet', which must be smaller than the oscillation length.
This is in agreement with the traditional explanation of these two
terms. Note that $L^{disp}_i\ll L^{disp}_j$ if $m_i\gg m_j$. In
that case, one can check that decoherence occurs for
$L<L^{disp}_i$.
\item
In the longitudinal-dispersion regime, the
coherence-length term mainly arises from the spreading of the
neutrino `wave packet': the interference term is averaged to zero
by the time integral as the spread of the wave packet becomes
large. The localization term mainly arises from the separation of
the `wave packets': if they are spatially separate at the onset of
the longitudinal-dispersion regime, they will never again overlap
because their separation increases as fast as their dispersion.
This is a new explanation for the origin of these two terms.
\end{itemize}
Thus the origins of the coherence-length and localization terms in
the transverse-dispersion regime become respectively the origins
of the localization and coherence-length terms in the
longitudinal-dispersion regime. The situation can be summarized in
the following diagram, where `wp' is an abbreviation for `wave packet':\\

\begin{tabular}{|c|c|c|}
\hline
   & transverse-dispersion  regime  & longitudinal-dispersion regime \\ \hline
Coherence length  & decreasing overlap of wp & increasing dispersion of wp \\
Localization      & initial spread of wp     & constant overlap of wp  \\
\hline
\end{tabular}\\
\begin{center}
TABLE I. Origins of the coherence-length and localization terms.
\end{center}

It is now possible to examine Shtanov's analysis in configuration
space \cite{shtanov}, since the explicit $L,T$ dependence of the
amplitude ${\cal A}_j$ has been computed for the whole range of
distances (see Eqs.~(\ref{nodispersion}), (\ref{transdisp}) and
(\ref{longdisp})). This author works directly with the neutrino
propagator in configuration space, and considers the space-time
variables $x,x'$ (see FIG.\ 1) as macroscopic variables. His model
will be considered in a scalar version in order to make the
comparison with our results easier. In configuration space, the
propagator for a scalar neutrino propagating from $x$ to $x'$ is
asymptotically ($m_j\sqrt{(x'-x)^2}\gg1$) given by
\begin{equation}
  {\cal A}_j\sim \frac{\sqrt{m_j}}{\left( (x'-x)^2 \right)^{3/4}} \,
  e^{-im_j\sqrt{(x'-x)^2} } \, ,
  \label{shtanovresult}
\end{equation}
where $(x'-x)^2$ is the Lorentz interval. Shtanov then computes
the convolution of this propagator with a source. In the case of a
monochromatic source, he obtains the standard oscillation formula.
However, a convolution with a strongly localized source
($\sigma_{x,t}\lesssim1/E$) leads to an amplitude that keeps its
dependence on the mass prefactor $\sqrt{m_j}$. In that case, the
transition probability is not equivalent to the oscillation
formula (\ref{oscformula}), at least if the neutrino masses are
not nearly degenerate. We proceed to show that Shtanov's result is
incorrect. Note first that the amplitude (\ref{shtanovresult}) is
in correspondence with our amplitude (\ref{longdisp}), computed
with the stationary phase method, since the prefactor in
Eq.~(\ref{longdisp}) can be rewritten, for $T \gg T^{disp}_j$, as
$$
  \frac{1}{T} \, \frac{\sigma_{peff}}{\sqrt{1+iT/T^{disp}_j}}
  \sim \frac{\sqrt{m_j}}{(T^2-L^2)^{3/4}} \, ,
$$
where the definition (\ref{dispersiontime}) of the dispersion time
has been used. This prefactor coincides with the prefactor in
Eq.~(\ref{shtanovresult}). However, the subsequent time average of
the transition probability completely cancels this dependence on
the mass, yielding Eq.~(\ref{oscformula}). This can be seen by
expanding the argument of the exponential in Eq.~(\ref{longdisp})
around the average propagation time. The width with respect to $T$
of the amplitude ${\cal A}_j$ is found to be equal to
$\frac{\sigma_{peff}T}{m_jv_0\gamma_{cl}^3}$ (where $\gamma_{cl}$
is the Lorentz factor associated with the velocity $v_{cl}=L/T$),
thereby providing a $m_j$-dependent factor that cancels the $m_j$
prefactor in the gaussian integration.

Shtanov does not obtain such a result, since he does not average
the probability on time. Besides, the mass prefactors would remain
even if a time average were performed on the probability: no `wave
packet envelope' appears indeed in Shtanov's amplitude, as this
author does not try to compute the explicit convolution of the
propagator (\ref{shtanovresult}) with a wave packet source of
arbitrary width. For that reason, he wrongly concludes that the
mass prefactor remains if the source is strongly localized.
Shtanov also derives another oscillation formula for mass
eigenstates such that $m_2\sqrt{(x'-x)^2}\ll1\ll
m_1\sqrt{(x'-x)^2}$. However, it can be checked that decoherence
occurs in that case. In conclusion, Shtanov's computations in
configuration space do not lead to new oscillation formulas.

The last case to be considered is the oscillation formula obtained
by Blasone, Vitiello and collaborators (hereafter abbreviated as
BV). These authors have attempted to define a Fock space of weak
eigenstates and to derive a nonperturbative oscillation formula
\cite{blasone}. They define flavor creation and annihilation
operators, satisfying canonical (anti)commutation relations, by
means of Bogoliubov transformations. Apart from the speculative
nature of the undertaking, the drawbacks of the approach are the
dependence on time, not on space, of the oscillation formula, as
well as the lack of observability conditions. The latter problem
is important since it determines whether the new features of the
BV oscillation formula are observable in practice. Since the BV
oscillation formula tends to the oscillation formula
(\ref{oscformula}) in the relativistic limit or if the mass
eigenstates are nearly degenerate (with the coherence-length and
localization terms removed), we can focus on the case of a
nonrelativistic oscillating particle having very distinct mass
eigenstates. In that case, $p\sim\delta m^2/2E$, so that either
$\sigma_p\lesssim\delta m^2/2E$ or $p\lesssim\sigma_p$. Under
these conditions, the oscillation formula (\ref{oscformula}) shows
that oscillations vanish, either because of the localization term
or because of the coherence-length term (though
Eq.~(\ref{oscformula}) is strictly speaking not valid in the case
considered, localization and coherence-length conditions are
generic features of quantum-mechanical models of oscillations
\cite{kayser}). Once the oscillation terms have been averaged to
zero both in the BV formula and in Eq.~(\ref{oscformula}), these
two oscillation formulas do not differ anymore. Therefore, the BV
formalism does not seem to be relevant to the phenomenology of
oscillations on macroscopic distances. This observation does not
detract from the theoretical worth of that approach.

We shall not treat here the charged lepton oscillations obtained
by Srivastava, Widom and Sassaroli \cite{srivastava}. The reader
is referred to \cite{beuthe} for a discussion of this subject.

\section{Conclusion}

In this article, we have shown that all correct results obtained
by applying quantum field theory to neutrino oscillations far from
the source are incorporated in the oscillation formula
(\ref{oscformula}). Closer to the source (i.e. for $L\lesssim
p_0/\sigma_p^2$), this formula is modified in only one respect
without repercussions on the oscillation length: the global
spatial distribution of the detection probability has the form
given in Eq.~(\ref{probanodisp}), with the consequence that the
neutrino propagation is restricted within a cylinder instead of a
cone. The quantum field oscillation formulas (\ref{oscformula})
and (\ref{probanodisp}) are equivalent to the standard result
(\ref{standard}) if some observability conditions are verified.
The new oscillations formulas proposed by Ioannisian and Pilaftsis
\cite{ioannisian} and by Shtanov \cite{shtanov} have been
disproved, while the Blasone-Vitiello formula \cite{blasone} has
been shown to be phenomenologically equivalent to the standard
result. Besides, arguments against oscillations of charged
leptons, as proposed by Srivastava, Widom and Sassaroli
\cite{srivastava}, can be found in Ref.~\cite{beuthe}.

We have insisted on the fact that the flavor-mixing transition
amplitude can be interpreted in terms of wave packets, so that
oscillations can be seen, like in the intermediate wave packet
model, as the result of an interference between propagating wave
packets. Moreover, the equivalence (\ref{incohsumQFT}) shows that
this physical picture still holds in the case of stationary
boundary conditions, provided that an incoherent sum over the
energy is performed, contrary to what was claimed in
Ref.~\cite{grimus99}. As a byproduct of our analysis, the
propagation of a particle has been shown to go through three
regimes, separated by two thresholds which are marked by the onset
of first the transversal dispersion and then the longitudinal
dispersion of the associated wave packet. This picture has been
found to be useful for the understanding of the terms present in
the oscillation formula as well as for the computation of the
correct spatial distribution of the probability. We believe that
the wave packet interpretation is the only consistent picture of
the oscillation process. In particular, it shows the irrelevance
of discussions bearing on the neutrino energy-momentum, since it
has not a unique value for a given mass eigenstate, but rather a
spread described by the overlap function $\psi(\sqrt{{\bf
p}^2+m^2_j},{\bf p})$.

One could wonder what advantage the external wave packet model has
over the intermediate wave packet model, apart from being the only
consistent way to derive the oscillation formula. Actually, the
quantum field formula (\ref{oscformula}) has two important new
features with respect to the quantum mechanical result. First, the
energy uncertainty is not put in by hand, but is defined in terms
of the 3-momentum widths and velocities of the in- and outgoing
wave packets (see Eq.~(\ref{energywidth})), with the result that
the stationary limit is well-defined and that it is possible to
analyze the dependence of the coherence length on the accuracy of
energy measurements. Secondly, the oscillation formula is valid as
well for neutrinos as for K and B mesons, putting it on a much
firmer phenomenological basis. This might be useful in the light
of the numerous nonstandard oscillation formulas existing in the
literature (see Ref.~\cite{beuthe} for a review).

The stationary limit has been considered in our computations to
show the great generality of the external wave packet model. The
stationary assumption is not problematic for the derivation of the
oscillation formula. However, this limiting case is rather
unphysical. The computations of section \ref{section4} have indeed
shown that boundary conditions can be considered as stationary
when $\sigma_{xeff}\gtrsim L$. This constraint, which can be
rewritten either as $\sigma_{e\SS P,D}\lesssim 1/L$ or as
$T^{overlap}_{\SS P,D}\gtrsim T$ (verified either at the source or
at the detector), is extremely stringent. In the example of
atmospheric neutrinos, the process can be considered as stationary
if $\sigma_{e\SS P,D}\lesssim10^{-19}\,$MeV or equivalently if
$T^{overlap}_{\SS P,D}\gtrsim10^{-3}\,$s. These conditions are far
from being realized at the microscopic level in any known neutrino
source or detector. Although most neutrino sources are stationary
from a macroscopic point of view, there is no reason to think that
individual particles in the source and detector remain unperturbed
in coherent states over macroscopic time scales: the Sun is
certainly not stationary at the atomic scale, and neither is a
detection process where charged leptons are observed with a finite
energy and time spread \cite{cardall}.

A subject deserving further investigation in the quantum field
framework is the influence of the decay width of the source of the
neutrino. The case of decay at rest has been studied in detail
with the Wigner-Weisskopf approximation in Ref.~\cite{grimus99},
while the treatment of decay in flight \cite{campagne} is still
unsatisfactory. Unstable neutrinos have been studied in
Ref.~\cite{beuthe}.

We would be sorry to leave the reader with the impression that the
quantum field approach to oscillations is complicated. Actually,
this is a consequence of the generality of the external wave
packet model. For pedagogical purposes, it is interesting to
resort to a simplified quantum field model, where the source and
detector are perfectly located in space and in- and outgoing
states are stationary \cite{kobzarev}.

\section*{Acknowledgments}

I thank Jeanne De Jaegher for helpful comments on the manuscript.


\footnotesize
\begin{thebibliography}{100}

\bibitem{bilenky}
S.~M.~Bilenky, C.~Giunti and W.~Grimus,
Prog.\ Part.\ Nucl.\ Phys.\ {\bf 43}, 1 (1999),
arXiv:hep-ph/9812360.

\bibitem{krastev}
J.~N.~Bahcall, M.~C.~Gonzalez-Garcia and C.~Pena-Garay,
JHEP 0108, 014 (2001), arXiv:hep-ph/0106258;
P.~I.~Krastev and A.~Yu.~Smirnov,
Phys.\ Rev.\ D {\bf 65}, 073022 (2002), arXiv:hep-ph/0108177;
V.~Barger, D.~Marfatia, K.~Whisnant and B.~P.~Wood,
arXiv:hep-ph/0204253.

\bibitem{kajita}
T.~Kajita and Y.~Totsuka,
Rev.\ Mod.\ Phys.\ {\bf 73}, 85 (2001).

\bibitem{K2K}
K2K Collaboration, S.~H.~Ahn {\it et al.},
Phys.\ Lett.\ B {\bf 511}, 178 (2001), arXiv:hep-ex/0103001.

\bibitem{lsnd}
LSND Collaboration, C.~Athanassopoulos {\it et al.},
Phys.\ Rev.\ Lett.\ {\bf 81}, 1774 (1998), arXiv:nucl-ex/9709006.

\bibitem{beuthe}
M.~Beuthe,
arXiv:hep-ph/0109119.

\bibitem{kayser}
B.~Kayser,
Phys.\ Rev.\ D {\bf 24}, 110 (1981);
C.~Giunti, C.~W.~Kim and U.~W.~Lee,
Phys.\ Rev.\ D {\bf 44}, 3635 (1991).

\bibitem{giunti92}
C.~Giunti, C.~W.~Kim,  U.~W.~Lee,
Phys.\ Rev.\ D {\bf 45}, 2414 (1992).

\bibitem{kiers96}
K.~Kiers, S.~Nussinov and N.~Weiss,
Phys.\ Rev.\ D {\bf 53}, 537 (1996), arXiv:hep-ph/9506271.

\bibitem{giunti98b}
C.~Giunti and C.~W.~Kim,
Phys.\ Rev.\ D {\bf 58}, 017301 (1998), arXiv:hep-ph/9711363.

\bibitem{giunti93}
C.~Giunti, C.~W.~Kim, J.~A.~Lee and U.~W.~Lee,
Phys.\ Rev.\ D {\bf 48}, 4310 (1993), arXiv:hep-ph/9305276.

\bibitem{giunti98}
C.~Giunti, C.~W.~Kim and U.~W.~Lee,
Phys.\ Lett.\ B {\bf 421}, 237 (1998), arXiv:hep-ph/9709494.

\bibitem{cardall}
C.~Y.~Cardall,
Phys.\ Rev.\ D {\bf 61}, 073006 (2000), arXiv:hep-ph/9909332.

\bibitem{kiers98}
K.~Kiers and N.~Weiss,
Phys.\ Rev.\ D {\bf 57}, 3091 (1998), arXiv:hep-ph/9710289.

\bibitem{kobzarev}
I.~Yu.~Kobzarev, B.~V.~Martemyanov, L.~B.~Okun and
M.~G.~Shchepkin,
Sov. J. Nucl. Phys. {\bf 35}, 708 (1982).

\bibitem{grimus96}
W.~Grimus and P.~Stockinger,
Phys.\ Rev.\ D {\bf 54}, 3414 (1996), arXiv:hep-ph/9603430.

\bibitem{grimus99}
W.~Grimus, S.~Mohanty and P.~Stockinger,
Phys.\ Rev.\ D {\bf 59}, 013011 (1999), arXiv:hep-ph/9807442;
{\bf 61}, 033001 (2000), arXiv:hep-ph/9904285.

\bibitem{ioannisian}
A.~Ioannisian and A.~Pilaftsis,
Phys.\ Rev.\ D {\bf 59}, 053003 (1999), arXiv:hep-ph/9809503.

\bibitem{chung}
C.~Y.~Cardall and D.~J.~H.~Chung,
Phys.\ Rev.\ D {\bf 60}, 073012 (1999), arXiv:hep-ph/9904291.

\bibitem{srivastava}
Y.~Srivastava, A.~Widom and E.~Sassaroli,
Eur.\ Phys.\ J.\ C {\bf 2}, 769 (1998).

\bibitem{shtanov}
Yu.~V.~Shtanov,
Phys.\ Rev.\ D {\bf 57}, 4418 (1998), arXiv:hep-ph/9706378.

\bibitem{blasone}
M.~Blasone and G.~Vitiello,
Ann.\ Phys.\ {\bf 244}, 283 (1995) [Erratum-ibid.\  {\bf 249}, 363
(1995)], arXiv:hep-ph/9501263;
M.~Blasone, P.~A.~Henning and G.~Vitiello,
Phys.\ Lett.\ B {\bf 451}, 140 (1999), arXiv:hep-th/9803157;
M.~Blasone and G.~Vitiello,
Phys.\ Rev.\ D {\bf 60}, 111302 (1999), arXiv:hep-ph/9907382;
M.~Blasone, P.~Jizba and G.~Vitiello,
Phys.\ Lett.\ B {\bf 517}, 471 (2001), arXiv:hep-th/0103087;
M.~Blasone, A.~Capolupo and G.~Vitiello,
arXiv:hep-th/0107125.

\bibitem{peskin}
M.~E.~Peskin and D.~V.~Schroeder, {\sl An Introduction to Quantum
Field Theory} (Addison-Wesley, Reading, 1995), p.~102.

\bibitem{jacob}
R.~Jacob and R.~G.~Sachs, Phys.\ Rev.\ {\bf 121}, 350 (1961);
R.~G.~Sachs, Ann.\ Phys.\ {\bf 22}, 239 (1963).

\bibitem{sudarsky}
D.~Sudarsky, E.~Fischbach, C.~Talmadge, S.~H.~Aronson and
H.-Y.~Cheng,
Ann.\ Phys.\ {\bf 207}, 103 (1991).

\bibitem{erdelyi}
A.~Erdelyi, {\sl Asymptotic Expansions} (Dover, New York, 1956);
C.~M.~Bender and S.~A.~Orszag, {\sl Advanced mathematical methods
for scientists and engineers} (McGraw-Hill, New York, 1978).

\bibitem{nussinov}
S.~Nussinov,
Phys.\ Rev.\ {\bf 63B}, 201 (1976).

\bibitem{campagne}
J.~E.~Campagne,
Phys.\ Lett.\ B {\bf 400}, 135 (1997).

\end{thebibliography}
\end{document}